\documentclass[bibyear]{aa}

\usepackage{times,txfonts}
\usepackage{graphicx,bm}
\newcommand\ab{\bm{a}}

\newcommand\rb{\bm{r}}

\newcommand\mub{\bm{\mu}}
\newcommand\sigmab{\bm{\sigma}}
\def\thetab{\bm{\theta}}

\def\url#1{\expandafter\string\csname #1\endcsname}

\def\lya{Ly$\alpha$}

\newcommand{\Lya}{\ifmmode{\mathrm{Ly}\alpha}\else Ly$\alpha$\xspace\fi}

\begin{document}

   \title{Three Lyman-$\alpha$ emitting filaments converging to a massive galaxy group at z=2.91: discussing the case for cold gas infall}
\titlerunning{Three Lyman-$\alpha$ emitting filaments converging to a massive galaxy group at z=2.91}

\author{E. Daddi\inst{1} \and F. Valentino\inst{2,3} \and R.~M.~Rich\inst{4} \and J.~D.~Neill\inst{5} \and M. Gronke\inst{6*} \and D.~O'Sullivan\inst{5}  \and
 D. Elbaz\inst{1} \and F. Bournaud\inst{1} \and \\   A. Finoguenov\inst{7} \and  A. Marchal\inst{8} \and I. Delvecchio\inst{1,9} \and S. Jin\inst{10,11} \and
 D. Liu\inst{12} \and V. Strazzullo\inst{13,14,15} \and A. Calabro\inst{16} \and R. Coogan\inst{17} \and \\C. D'Eugenio\inst{1} \and R. Gobat\inst{18} \and  B.~S.~Kalita\inst{1} \and
  P. Laursen\inst{19,2} \and D.C. Martin\inst{5} \and A. Puglisi\inst{20} \and E. Schinnerer\inst{12}  \and T. Wang\inst{21}
     } 

\offprints{Emanuele Daddi
\email{edaddi@cea.fr}}

\institute{
CEA, Irfu, DAp, AIM, Universit\`e Paris-Saclay, Universit\`e de Paris, CNRS, F-91191 Gif-sur-Yvette, France
\and Cosmic Dawn Center (DAWN) 
\and Niels Bohr Institute, University of Copenhagen, Jagtvej 128, DK-2200, Copenhagen N, Denmark
\and Department of Physics \& Astronomy, University of California Los Angeles, 430 Portola Plaza, Los Angeles, CA 90095, USA
\and California Institute of Technology, 1216 East California Boulevard, Pasadena, California 91125, USA.
\and Department of Physics \& Astronomy, Johns Hopkins University, Baltimore, MD 21218, USA\thanks{Hubble Fellow}
\and Department of Physics, University of Helsinki, Gustaf H{\"a}llstr{\"o}min katu 2, FI-00014 Helsinki, Finland
\and Canadian Institute for Theoretical Astrophysics, University of Toronto, 60 St. George Street, Toronto, ON M5S 3H8, Canada.
\and INAF - Osservatorio Astronomico di Brera, via Brera 28, I-20121, Milano, Italy \& via Bianchi 46, I-23807, Merate, Italy
\and Instituto de Astrof\'isica de Canarias (IAC), E-38205 La Laguna, Tenerife, Spain 
\and Universidad de La Laguna, Dpto. Astrof\'isica, E-38206 La Laguna, Tenerife, Spain
\and Max Planck Institute for Astronomy, Konigstuhl 17, D-69117 Heidelberg, Germany
\and Faculty of Physics, Ludwig-Maximilians-Universit\"at, Scheinerstr. 1, 81679, Munich, Germany
\and Dipartimento di Fisica, Universit\'a di Trieste, Via Tiepolo 11, I-34143 Trieste, Italy
\and INAF-Osservatorio Astronomico di Brera, via Brera 28, 20121, Milano, Italy
\and INAF, Osservatorio Astronomico di Roma, via Frascati 33, 00078, Monteporzio Catone, Italy
\and Max-Planck-Institut fur extraterrestrische Physik (MPE), Giessenbachstr. 1, 85748 Garching, Germany
\and Instituto de Fisica, Pontificia Universidad Catolica de Valparaiso, Casilla, 4059, Valparaiso, Chile
\and Institute of Theoretical Astrophysics, University of Oslo, PO Box 1029 Blindern, 0315 Oslo, Norway
\and Centre for Extragalactic Astronomy, Durham University, South Road, Durham DH1 3LE, UK
\and Institute of Astronomy, The University of Tokyo, 2-21-1 Osawa, Mitaka, Tokyo 181-0015, Japan
 } 

\date{Received / Accepted }

\abstract{
We have discovered a 300~kpc-wide giant \lya\ nebula centered on the  massive galaxy group RO-1001 at z=2.91 in the COSMOS field.  Keck Cosmic Web Imager observations reveal three cold gas filaments converging into the center of the potential well of its $\sim 4\times10^{13}$~M$_\odot$  dark matter halo, hosting 1200~M$_\odot$~yr$^{-1}$ of star formation as probed by ALMA and NOEMA observations. The nebula morphological and kinematics properties and the prevalence of blueshifted components in the \lya\ spectra are consistent with a scenario of gas accretion. The upper limits on AGN activity and overall energetics favor gravity as the primary \lya\ powering source and infall as the main source of gas flows to the system.  Although interpretational difficulties remain, with outflows and likely also photoionization with ensuing recombination still playing a role, this finding provides arguably an ideal environment to quantitatively test models of cold gas accretion and galaxy feeding inside an actively star-forming massive halo at high redshift.
} 

\maketitle

%

\section{Introduction}

A fundamental phenomenon required to explain the evolution of massive galaxies at high redshifts  is the efficient accretion of cold gas streaming along filaments, surviving the shocks at the virial radii of their massive halos and delivering the required fuel to galaxies (Dekel et al. 2009; Kere{\v s} et al. 2005). This scenario is intimately connected to our current understanding of the star formation and growth of galaxies at {\it cosmic noon} $1<z<3$ (and earlier), whose key observational features might be summarized with two basic tenets: the existence of tight correlations between the stellar mass and star formation rates (SFRs) in galaxies (the so-called Main Sequence of star formation; e.g., Noeske et al. 2007; Elbaz et al. 2007; Daddi et al. 2007) and the systematic increase of gas fractions along with specific SFRs as a function of redshift for typical Main Sequence galaxies (e.g., Daddi et al. 2008; 2010; Tacconi et al. 2010; Magdis et al. 2012; Genzel et al. 2015). The finding that star forming galaxies at these redshifts are much more common than quiescent systems (e.g., Ilbert et al. 2010), coupled to the tight Main Sequence correlations imply that star formation in galaxies occurs  and is persistent over timescales much longer than their typical stellar doubling times and gas consumption timescales, which requires constant replenishment of their gas reservoirs (e.g., Lilly et al. 2013; Walter et al. 2020). 

Cold accretion models quite satisfactorily account for this observational evidence, as they predict that cold material, nearly ready to form stars, accretes at rates proportional to the hosting halo mass (Neinstein \& Dekel 2008; Dekel et al. 2013), thus naturally resulting in Main-Sequence like behavior (as recognized by theory even before observational confirmation, see e.g. Finlator et al. 2006). Also, accretion rates at fixed mass are predicted to evolve rapidly with redshift, with trends  (scaling as $(1+z)^{\alpha}$ with $\alpha\approx2$--3)  that correspond well to the evolving behavior of the Main Sequence normalization (e.g., Sargent et al. 2012) and gas fractions (Magdis et al. 2012; Genzel et al. 2015). { It is crucial though that accreting gas be cold -- hot accretion occurs on too long timescales and at too low rates to be effective (Keres et al. 2005; Katz et al. 2002; Birnboim \& Dekel 2003). }
Still, even in the assumption of cold accretion not everything is fully reconciled. For example a tension between predicted versus observed star formation rates in typical galaxies at {\it cosmic noon} has persisted for over a decade (e.g., Daddi et al. 2007) but it is generally understood as due to limitations in the modelling of feedback and the subsequent implications for  gas consumption and the baryon cycle (Somerville \& Dave 2015; Popping et al. 2019).

Now, despite more than a decade of effort,  definitive observational confirmation of the existence of such cold  gas accreting from the intergalactic medium are still lacking, so that the theory is being necessarily questioned. { On the observational side it appears that outflows are actually widespread in absorption in the circumgalactic gas around galaxies (Steidel et al. 2010), making it harder to study inflowing gas in this way. A number of studies at moderate redshifts report evidence of enriched gas inflows from redshifted components of metal lines in absorption (Giavalisco et al. 2011; Rubin et al. 2012; Martin et al. 2012; Bouche et al. 2013; 2016; Turner et al. 2017; Zabl et al. 2019; Chen et al. 2020; Fu et al. 2021). It remains often unclear, however, if this is cold accretion rather than gas recycling or infall connected with mergers.
From the theoretical side, the latest generation, high resolution simulations now call into question whether streams can  survive the interaction with the hot baryons in halos and remain stable (Nelson et al. 2015; Mandelker et al. 2019). Also, numerical simulations of cold streams have been questioned for not having the required resolution to capture the small scale gas physics (Cornuault et al. 2018), making it unclear whether predictions can be taken quantitatively. 
This uncertainty on the feeding of galaxy activity also limits our understanding of  feedback processes (e.g., Gabor \& Bournaud 2014; Dekel \& Mandelker 2014).

It is widely recognized that the most promising avenue to reveal these cold gas streams is through their collisionally  excited \lya\ emission (Dijkstra \& Loeb 2009;  Goerdt et al. 2010; Rosdahl \& Blaizot 2012), possibly enhanced by hydrodynamical instabilities (e.g., Mandelker et al. 2020a). Much more difficult is to ascertain whether any observed extended \lya\ emission  is due to collisional excitation or rather than recombinations following photo-ionization from star formation and/or AGN activity. Even more fundamental is the difficulty of properly distinguishing \lya\ emission as arising from outflowing versus infalling gas, given that broadly, either mass motion would give rise to similar instability-driven phenomenology (e.g., Cornuault et al. 2018; Qiu et al. 2020). 

Giant \lya\ nebulae are now routinely discovered around QSOs at redshifts $2<z<4$ (e.g., Borisova et al. 2016; Arrigoni Battaia et al. 2018; 2019; Cai et al. 2019; O'Sullivan et al. 2020)  with detections as high as $z\sim6.6$ (Farina et al. 2019) and could potentially provide large samples to statistically search for the role of infall. 
Filamentary structures, sometimes found in QSO \lya\ nebulae (Cantalupo et al. 2014; Hennawi et al. 2015), might be consistent with gas infall (e.g., Martin et al. 2015a; 2019). However, it is not easy to rule out  alternative interpretations:  outflows (Fiore et al. 2017; Guo et al. 2020; Veilleux et al. 2020) overshadow expected infall  in luminous QSO hosting halos by orders of magnitudes for both energy and gas flows (see quantitative discussion later in this work; Fu et al. 2021). Also, the \lya\ emission in those environments is almost certainly photoionized by the QSO hard UV emerging photons, making it prohibitive not only to gauge if any gravitational driven \lya\ is at all present in QSO nebulae but also if any infall is actually taking place. 

{ A remarkable Mpc-long filament shining in \lya\ has been recently found also in the  SSA22a-LAB1 protocluster environment at $z=3.1$ (Umehata et al. 2019). That relatively giant \lya\ nebula, tough, does not appear to be consistent with arising from infall alone, 
as there is a large shearing velocity field orthogonal to the main axis of the nebula (Herenz et al. 2020).  There is no direct kinematic evidence of gas accretion or convergence onto the deep potential well of any particular massive dark matter halos in the Mpc-scale filaments reported in SSA22, likely due to the sparsity and extension of this young proto-cluster region encompassing a number of active galaxies. However, this could be the case on individual sub-regions and smaller blobs. 
The filamentary structures around individual QSOs and in the SSA22 protocluster were suggested as signatures of the connection with the cosmic web, where each individual star-forming galaxy or QSOs locally illuminates their surrounding gas via photoionization, enabling the detection.
However, such filaments are extended over scales much longer than the virial radius or any meaningful size metrics of their putative hosting dark matter halos, which is where theoretical works have predicted the possible detection of Lya emission from cold flows owing to sufficient gas density and confinement from hot gas (Dekel et al. 2009, Dijkstra \& Loeb 2009). The nature and origin of these filaments is thus still an interesting and open problem to date.}

A critical test for models would then be the search for cold accreting gas in distant and  massive halos and in environments where the  contrast with competing mechanisms for gas flows and for powering the detectable Lya emission is maximal. The first requirement follows from the fact that the dark and baryonic 
matter accretion rates increase with both the halo mass and redshift (Neinstein \& Dekel 2008, 
Dekel et al. 2009), and is not trivial to address when considering that massive halos become more rare in
the distant Universe because of their hierarchical assembly. Moreover, the necessity of excluding
alternative mechanisms suggests a move away from extreme sources such as QSOs, focusing on structures where the black hole and
star formation activities proceed at a standard pace. Both lines of argument point to high redshift clusters or groups as ideal testbeds for comparing theory to observations and searching for the evidence of cold accreting gas, as already seminally suggested in Valentino et al. (2015; see their Fig.~17 and related discussion) and Overzier et al. (2016; see their Fig.~11 and related discussion).
 This is because such clusters/group would provide the opportunity to search for non-photoionized \lya\ in an environment where the role of outflows could be minimal and where filaments could be studied in connection to the halo they are streaming into, thus enabling quantitative comparison to cold accretion theory.
 This work presents one such plausible candidate. { A more complete discussion comparing to prospects for detecting cold accretion in lower mass halos is postponed to the conclusion section, as it will benefit from the presentation of our observational results.}
 
Following the serendipitous discovery (Valentino et al. 2016) of a giant \lya\ halo centered on the X-ray detected cluster CL~1449 at $z=1.99$, we have pursued this avenue and started systematic observations of several structures at $2<z<3.5$ with the Keck Cosmic Web Imager (KCWI), searching for redshifted \lya. This is reversing the standard strategy of discovering \lya\ nebulae from blind (e.g., narrow-band) surveys and following them up to find that they are typically hosted in moderately dense environments, by starting with a systematic investigation of the prevalence of \lya\ emission inside massive halos at high redshifts.  Among the great advantages of integral field spectroscopy, as now provided routinely by MUSE (Bacon et al. 2010) and KCWI (Morrissey et al. 2018), and as opposed to earlier  narrow-band imaging attempts, is the potential to unveil the kinematics and spectral properties of the \lya\ emission that,  keeping in mind the uncertainties linked to resonant scattering effects, can provide valuable diagnostics on the presence of accretion (see, e.g., Ao et al. 2020). 
As part of these efforts, 
 we used KCWI   to search for redshifted \lya\ in RO-1001, a massive group of galaxies at $z=2.91$ that currently is our best studied target with the deepest and widest observations, that we present in this work. Results for our full KCWI survey of distant structures will be presented elsewhere (E. Daddi et al., in preparation).
 
 This paper is organized as follows: we present in Section~2 the observational characterization of the RO-1001 structure, starting from the observations of the giant \lya\ halo which motivates a detailed look into the overall properties of galaxies hosted thereby.  Sect.~3 presents the spectral properties of the \lya\ emission including moments (velocity and dispersion fields).  We interpret them with the aid of simplified resonant scattering modeling and multi-Gaussian decomposition, and compared to cold accretion theory predictions informed by the estimated DM halo.
Sect.~4 discusses the overall energetics and gas flows that characterize the system, particularly in comparison with \lya\ nebulae observed around QSOs. Conclusions and summary are provided in Section~5. 
 In this work we adopt a standard cosmology and a Chabrier IMF.  

\begin{figure*}[t]
{\centering  \includegraphics[width=21cm]{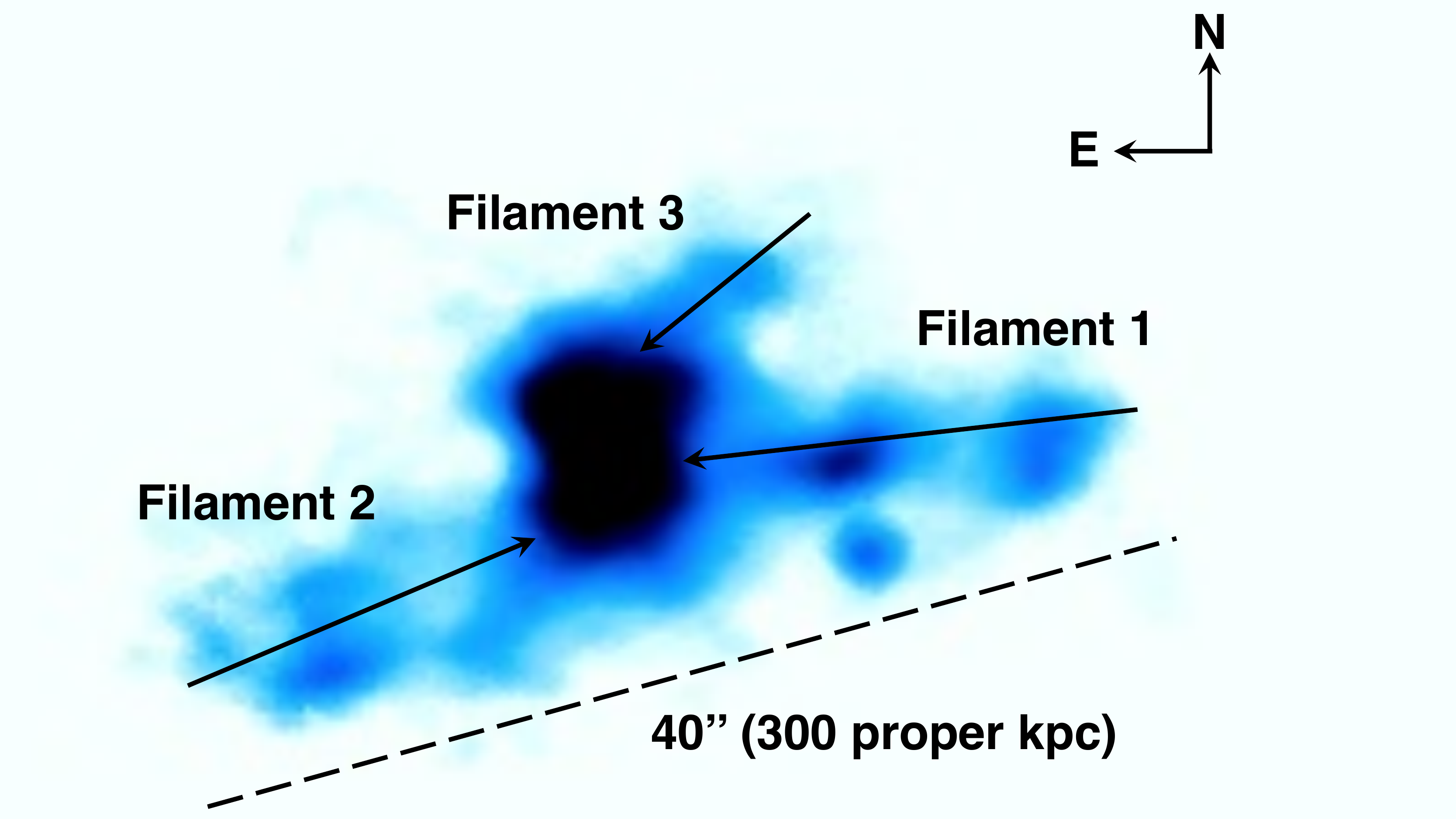}}
\caption{
Ly$\alpha$ image from KCWI observations of the RO-1001 group. The three extended filaments that are clearly traced by \lya\ are labeled. The corresponding \lya\ surface brightness levels can be gauged from Fig.~2. 
}
\end{figure*}

\begin{figure*}[t]
{\centering  \includegraphics[width=\textwidth]{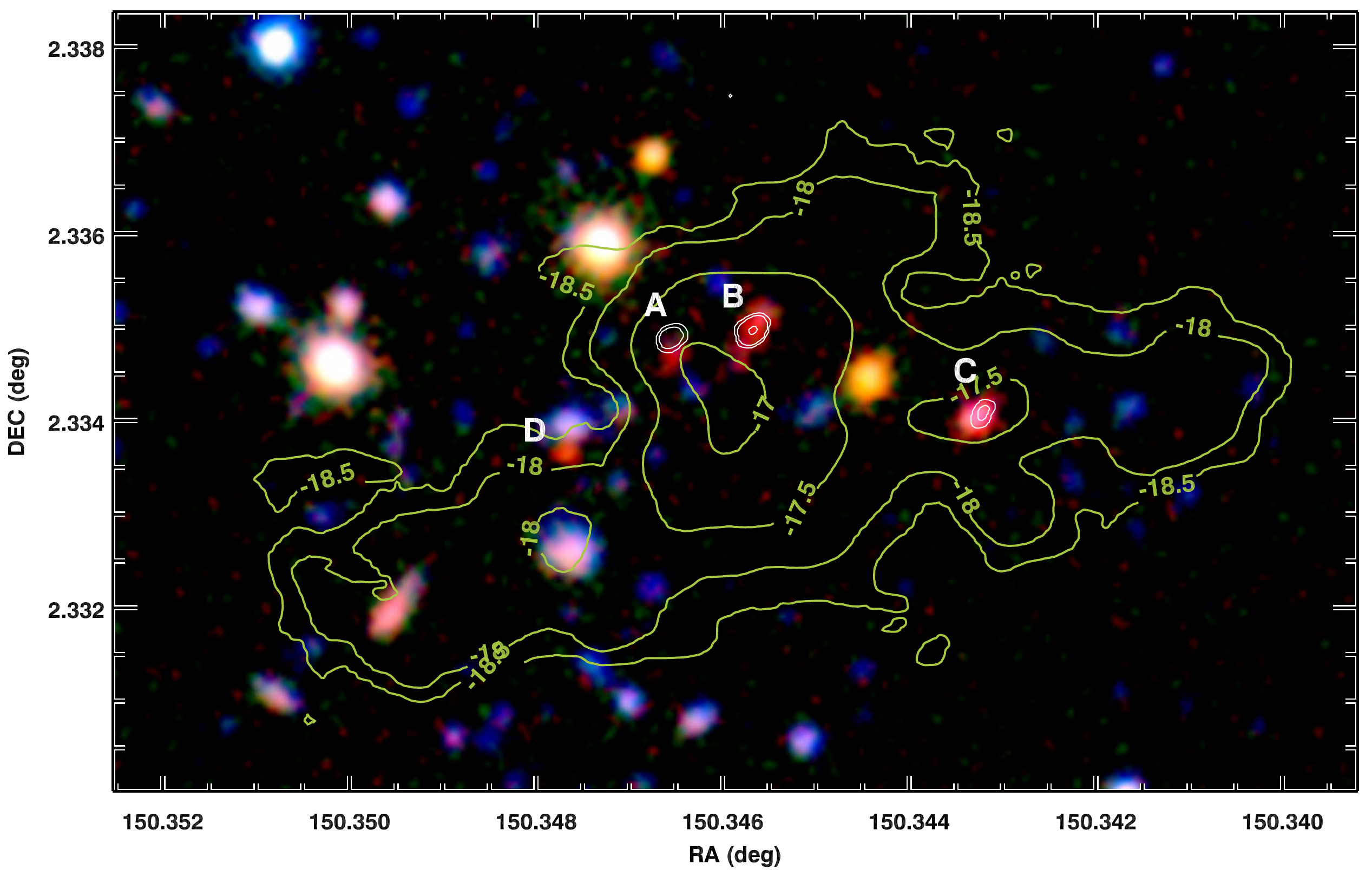}}
\caption{
Deep optical-near-infrared imaging of the RO-1001 group from UltraVISTA Ks and J (red and green) and Subaru Suprime CAM (combined VRI bands; blue), over a $30''\times40''$ field of view. The green contours show the \lya\ surface brightness  (SB)  from Figure~1 in steps of 0.5~dex, from a few $10^{-17}$~s$^{-1}$~erg~cm$^{-2}$~arcsec$^{-2}$ (center) to $\times30$ lower surface brightness (filaments). 
The three ALMA/VLA sources are labeled (ABC; white contours), together with a fourth massive galaxy (D) that is a candidate quiescent object in the structure. Notice that further luminous objects aligned with the \lya\ emission are in the foreground, see e.g. the \lya\ depression coincident with the two sources in the East filament, that we interpret as due to dust absorption. 
}
\end{figure*}

\section{Observational characterization of  RO-1001 }

RO-1001 was selected in the COSMOS 2-square-degree field as a 12$\sigma$ overdensity of optically faint radio sources centered at RA 10:01:23.064  and DEC 2:20:04.86, following a  recently proposed technique  (Daddi et al. 2017): it  was found to contain 3 VLA detections with
$S_{\rm 3\,GHz}>30\,\mu$Jy and $z_{\rm phot}>2.5$ within a radius of 10$''$ (80 kpc; proper scales are used throughout the paper), i.e. the size of a massive halo core (Strazzullo et al. 2013; Wang et al. 2016). 
These three VLA galaxies are also bright in ALMA sub-mm imaging  ($200\,\mu$m rest-frame), i.e. they are highly star-forming galaxies   (not AGNs), as discussed in the following. NOEMA CO[3-2] line observations confirm their $z_{\rm spec}\sim2.91$. In this section we first present the observations and reduction of KCWI data, with the discovery of a \lya\ nebula in this structure. We then discuss the stellar mass of its member galaxies, and constrain the hosting halo mass. Finally, we present the exploration of their star formation and AGN content and the derivation of redshifts from CO observations. 

\subsection{KCWI observations and analysis.}

We observed RO-1001 with KCWI on January 16th 2018 for 1~h using the BM grism ($R=2000$ with the adopted large field of view), and on February 3rd and 4th 2019 for 3.5~h and 4~h, respectively, using the lower resolution BL grism ($R=900$), giving a total 8.5~h on-source when combining all observations. Conditions were excellent with dark sky and seeing typically in the range of 0.4--0.7$''$. We used integration times of 900~s in 2018 and of 1800~s in 2019, with dithering and large offsets to eventually cover a region corresponding to 2$\times$2 KCWI fields of view of $40''\times60''$ (i.e. $300\times500$~kpc$^2$). In fact, already with the first 1~h integration obtained in 2018 it was clear that the nebula  extends well beyond the usable KCWI field of view of about $18''\times31''$ for the adopted configuration with large slices ($1.35''$). The low-resolution BL data covers the full 3500--5500\AA\ range  (corresponding to 900--1400~\AA\ rest-frame at the redshift of RO-1001), while the BM observations cover a shorter range across the \lya\ emission.  We used the standard KCWI pipeline {\it Kderp} (Morrissey et al. 2018; Neill et al. 2018)  for  the data reduction, including twilight flats to obtain accurate illumination corrections.  We further used  {\it CubEx} tools (Cantalupo et al. 2019) to refine the flat-fielding slice by slice, thus allowing us to improve the sky subtraction by removing a median sky value at each wavelength layer, after masking sources detected in the stacked (continuum) cubes. This step was first performed over the whole frame and subsequently iterated by masking regions where \lya\ emission had been detected, to avoid self-subtraction of the \lya\ signal.   
Further reduction and analysis steps were performed with the {\em CWItools} scripts (O'Sullivan et al. 2020; Martin et al. 2019). We estimate variance cubes from the original, non-resampled cubes and propagate the uncertainties through the combination to obtain a final variance cube. We combined the dithered and offset observations based on the 
astrometry of each frame that was derived by cross-correlating to B-band Suprime CAM imaging of the area publicly available from the COSMOS survey, and resampling into a final pixel scale of 0.29$''$, corresponding to the finer grid in the original slices.
 We then subtracted any continuum emission from objects in the combined cubes by fitting a 7th order polynomial as a function of wavelength at each spatial pixel, avoiding the wavelength range where \lya\ emission is present. From compact objects in the final cube we estimate an average image quality of 0.6$''$ (FWHM). The \lya\ nebula is very clearly detected (Fig.~1 and~2). We produced a low-resolution cube containing all 8.5~h observations, used for most of the analysis in this paper, and a higher spectral resolution cube using only 2018 BM data, used to obtain higher quality \lya\ spectra in the core (Sect.~3). 
We used   adaptive smoothing (O'Sullivan et al. 2020; Martin et al. 2019)  to recover the full extent of the nebula in the  low-resolution cube, thresholding at the 3$\sigma$ level. We started by smoothing with a spatial kernel equal to the seeing and averaging three (1\AA\ wide) spectral layers (roughly corresponding to the spectral resolution). This already selects 93\% of the pixels eventually detected in the \lya\ nebula when allowing for larger spatial and spectral smoothing scales. { The SNR of the \lya\ emission in the original, unsmoothed data
is shown in Fig.~A1.} 

The observations revealed faint low-surface brightness, large area \lya\
filamentary structures that converge onto a  bright \lya\ nebula  at the center of the potential well of the group  (Figs.1,~2),  
with a total  
$L_{Lya}=1.3\pm0.2\times10^{44}$~erg~s$^{-1}$ (the uncertainty includes the effect of SNR thresholding on flux detection and accounting for possible absorption\footnote{See Laursen et al. 2011; Dijkstra \& Loeb 2009.}  from $z=2.91$). 
Three filaments can be readily recognized from the surface brightness profile of the nebula (Figs.1,~2). The two most prominent, extending South-East and West of the nebula, respectively,  appear to be traceable over a projected distance of  200~kpc from the core, at the current 3$\sigma$ surface brightness limit of $10^{-19}$~erg~s$^{-1}$~cm$^{-2}$~arcsec$^{-2}$, reconstructed from adaptive smoothing. A third structure that we can identify with a filament extends towards the North-West (Figs.~1,2), likely affected by projection effects. 

{ Moment~1 (velocity) and~2 (dispersion) maps are built using pixels flagged to be part of the nebula's detected signal by the  adaptive smoothing  procedure (as discussed above), but using original (unsmoothed) pixel values. Uncertainties in the moment maps are obtained by error propagation. }

 \begin{figure}
{\centering  \includegraphics[width=9cm]{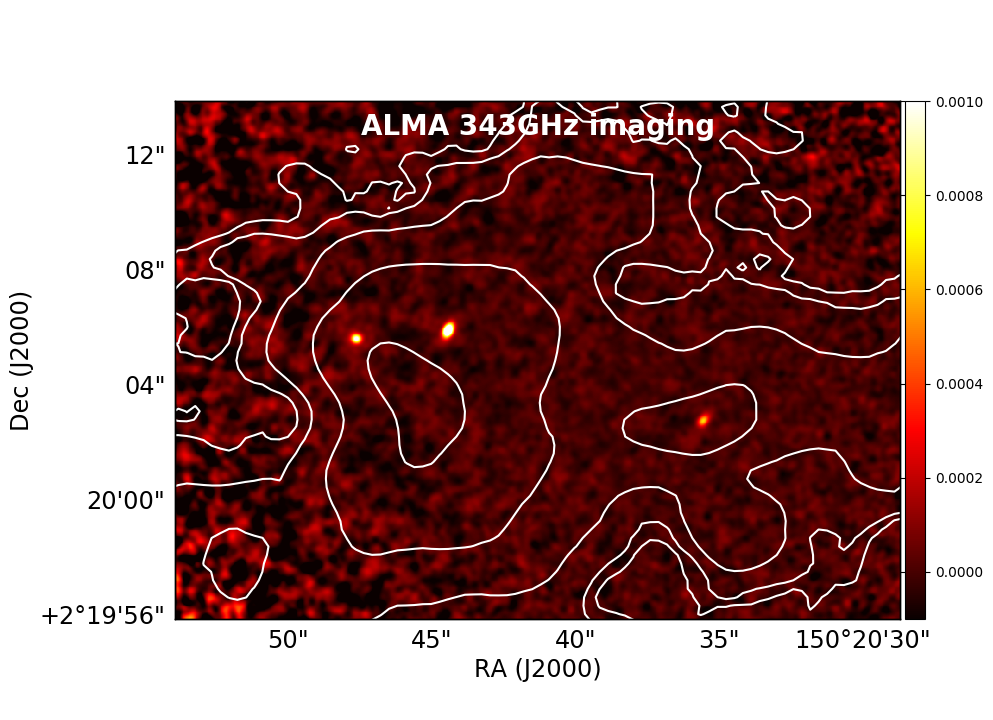}}
\caption{ALMA  343~GHz  continuum mosaic image with primary beam attenuation correction (the colorbar shows Jy~beam$^{-1}$). The image reaches an rms of $\sim$50\,$\mu$Jy\,beam$^{-1}$ at the center, at the resolution of 0.15$''$.   Contours show the \lya\ emission.
}
\end{figure}

\begin{figure*}
{\centering  
\includegraphics[width=0.99\textwidth]{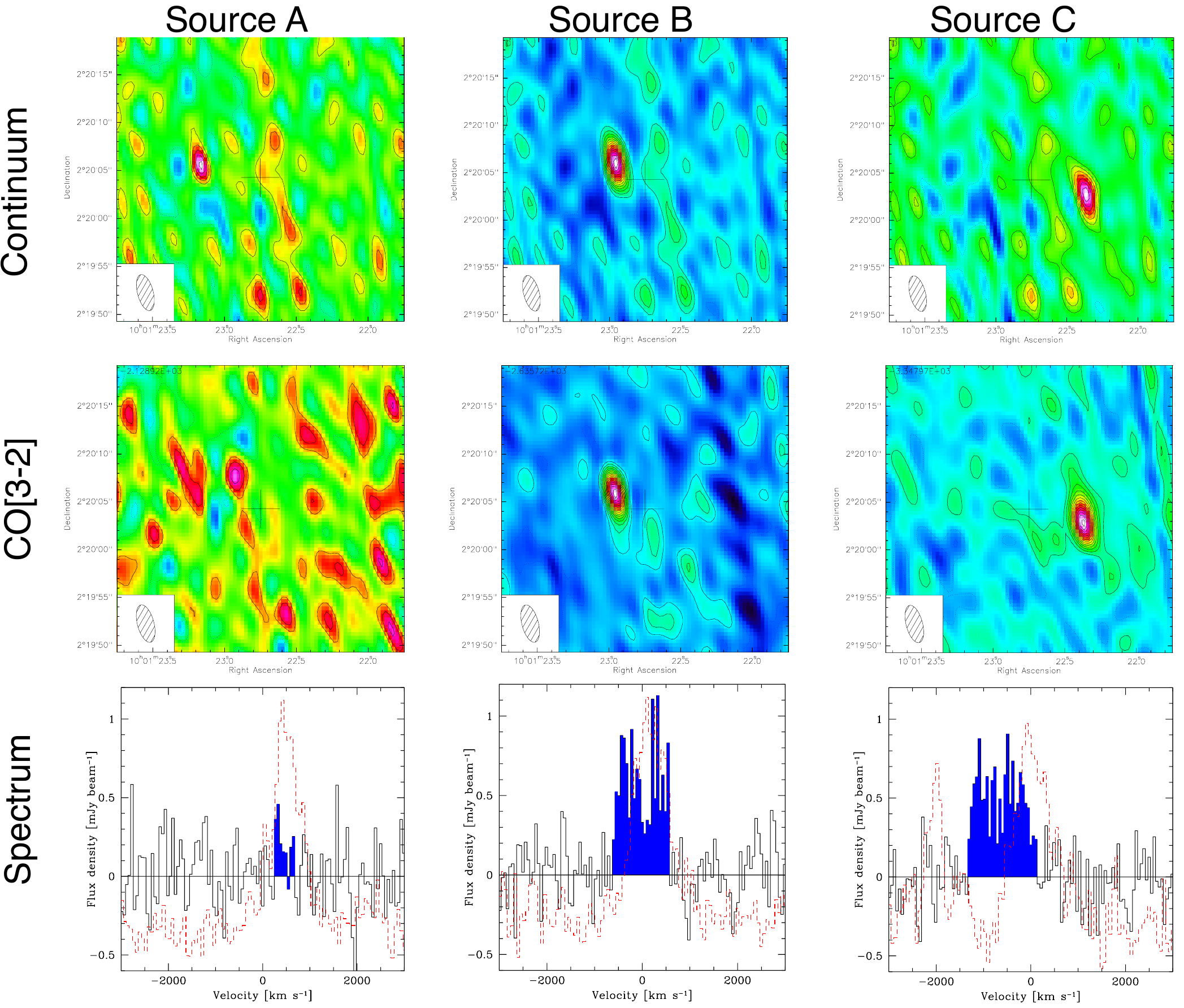}
}
\caption{
NOEMA observations. The top row shows continuum images, the second row integrated CO[3-2] line emission, and the third one the spectra. 
Different columns correspond to sources A, B, C, as labeled.
 The NOEMA spectrum of source A, extracted at the well determined continuum position 
(SNR$=8$) shows a faint 3.5$\sigma$ line, at the  velocity of the \lya\  line at the same position {  The \lya\ spectra extracted over a PSF at the same location of the galaxies are shown as red dashed histograms, shifted in the plots to lower zero-level for easier comparison. The \lya\ spectra are from the lower-resolution, higher SNR KCWI cube.} The cross
in each map is $\pm2''$ and marks the position of the phase center of the NOEMA observations.
}
\end{figure*}

\subsection{Stellar masses and hosting halo mass.}

There are four massive ($M>10^{11}\,M_\odot$) galaxies within $13''$ ($\sim$100 kpc proper at $z=2.91$) of RO-1001, with an
estimated (Muzzin et al. 2013; Laigle et al. 2016) photometric redshift in the range $2.5<z<3.5$, see Table~1.
One source  (D in Fig.~2) is blended
with a close neighbor. We
obtained a revised stellar mass estimate, empirically calibrated on
sources at similar redshift from the COSMOS2015 catalog (Laigle et al. 2016), and
based on J, H, K photometry from the DR4 UltraVISTA imaging and 3.6,
4.5$\mu$m Spitzer/IRAC imaging. 
We estimate a stellar mass completeness limit  of log(M/M$_{\odot}$)=10.8. 

Three of the massive galaxies are shown to be at $z=2.91$ from CO[3-2] spectroscopy, as discussed later in this section,
but source D remains unidentified as it is likely passive (Sect.~2.3).  Given the  similarity in the optical rest-frame colors and SED to other group members, and the negligible probability of such a red and massive galaxy being  there by chance (recall galaxy D was not detected in the radio),  we will assume in the following that it also resides in the RO-1001 group.

There are no additional $z\sim3$ massive $\sim10^{11}M_\odot$ galaxies to
a distance of $1'$ from RO-1001. 
The total stellar mass of $\log(M/M_\odot)>10.8$ galaxies in the structure thus adds up to $5.4^{+2}_{-0.5} \times 10^{11}\,M_\odot$, where we assumed that the uncertainty on individual stellar mass estimates is at least a factor $\sim$50\%, see e.g. Muzzin et al. 2013. We extrapolate a total stellar mass down to $10^{7}\,M_\odot$ of
1.0$^{+0.7}_{-0.2} \times10^{12}$M$_{\odot}$ assuming the  stellar mass function of field galaxies, again from Muzzin et al. 2013,
  at $2.5<z<3$. Adopting the scaling between total stellar and halo mass derived  from $z\sim1$ clusters with masses in the range 0.6--16 $\times 10^{14}\,M_\odot$ (van der Burg et al. 2014),  would yield a halo mass of $M_{200} \sim 6 \times 10^{13}\,M_\odot$.
  We note that if scaling by the difference between the cluster and field galaxy stellar mass function at $z\sim 1$ as in van der Burg et al. 2013, we would obtain a lower estimate for the total stellar mass of $7.6^{+5}_{-1.5} \times10^{11}$M$_{\odot}$, and thus a lower estimate for the total halo mass of
$M_{200}\sim4\times10^{13}\,M_\odot$. However, the environmental dependence of stellar mass functions at $z\sim1$ and $z\sim3$ may be significantly different, so a range of $M_{200}\sim4$--$6\times10^{13}\,M_\odot$ brackets  plausible estimates. We note that the stellar mass-to-DM mass scaling that we obtain is similar to what is estimated for Cl-1001 and for Cl-1449, which are supported by X-ray detections (Gobat et al. 2011; Wang et al. 2016) and SZ for Cl-1449 (Gobat et al. 2019).

\begin{table*}
{
\caption{Massive galaxies in the RO-1001 group}
\label{tab:1}
\centering
\begin{tabular}{clllll}
\hline\hline
ID                        &                       & A             & B              & C                 & D \\
\hline
RA                        &                       & 10:01:23.174  & 10:01:22.964   & 10:01:22.369      & 10:01:23.438 \\
DEC                       &                       & 02:20:05.57   & 02:20:05.87    & 02:20:02.63       & 02:20:01.10 \\
$z_{\rm spec}$            &                       & 2.9214        & 2.9156         & 2.9064            & 2.9 (1) \\
$\log M_{\star}$          & $M_\odot$             & 11.13         & 11.13          & 11.23             & 11.00 \\
SFR (2)                   & $M_\odot$~yr$^{-1}$   & 306           & 706            & 266               & $<30$ \\
$S_\nu$(870~$\mu$m)       & mJy                   & $4.0\pm0.1$   & $9.1\pm0.1$    & $3.39\pm0.15$     & $<0.3$ \\
$S_\nu$(1.25~mm)          & mJy                   & $1.2\pm0.1$   & $3.4\pm0.1$    & $1.3\pm0.2$       & $<0.3$ \\
$S_\nu$(3.4~mm)           & $\mu$Jy               & $40\pm5$      & $88\pm6$       & $39\pm5$          & $<15$ \\
$S_\nu$(10~cm)            & $\mu$Jy               & $38\pm3$      & $34\pm3$       & $69\pm6$          & \\
I$_{\rm CO[3-2]}$         & Jy$\times$km~s$^{-1}$ & $0.10\pm0.03$ & $0.69\pm0.05$  & $0.63\pm0.05$     & $<0.1$ (3) \\
FWZV$_{\rm CO[3-2]}$ (4)  & km~s$^{-1}$           & 381           & 1114           & 1098 \\
$v_{\rm CO[3-2]}$ (5)     & km~s$^{-1}$           & 460           & 13             & -690 \\
$r_{\rm 1/2}$  (6)        & $''$                  & $0.07\pm0.01$ & $0.10\pm0.003$ & $0.11\pm0.007$ \\
\hline
\hline
\end{tabular}\\
}
{Notes: (1) photometric redshifts; (2) derived from the measurement of individual galaxies assuming the same SED shape as for their coaddition (Fig.~5). (3) assuming a linewidth of 500~km~s$^{-1}$.  (4) Full Width at Zero Velocity corresponding to the full extraction range of the emission line in velocity.  (5) systemic velocities of the galaxies are relative to the average, flux-weighted redshift of the \lya\ emission ($z=2.9154$). (6) reported sizes  are half-light radii from a circular Gaussian fit. Errors are much smaller than the beam size given the high SNR detections. The average size of $0.1''$ corresponds to 800~pc at $z=3$. 
}
\end{table*}

\begin{table}
{
\caption{Integrated IR and radio emission in the RO-1001 group (see also Fig.5)}
\label{tab:X}
\centering
\begin{tabular}{cll}
\hline\hline
IRAC ch1                         & $45.85 \pm 0.26 $  & $\mu$Jy       \\
IRAC ch2                         & $61.6 \pm 0.3 $  & $\mu$Jy       \\
IRAC ch3                         & $69 \pm 6 $   & $\mu$Jy       \\
IRAC ch4                         & $71 \pm 8 $     & $\mu$Jy       \\
MIPS 24$\mu$m                           & $166\pm21$    & $\mu$Jy       \\
PACS 100$\mu$m                         & $2.42 \pm 1.66$    & mJy           \\
PACS 160$\mu$m                         & $0.3 \pm 3.2$   & mJy           \\
SPIRE 250$\mu$m            & $27.5\pm1.8$    & mJy           \\
SPIRE 350$\mu$m                         & $42.2\pm2.5$   & mJy           \\
SPIRE 500$\mu$m                         & $45.7\pm3.6$    & mJy           \\
SCUBA 850$\mu$m                    & $19.2\pm1.7$     & mJy           \\
ALMA 870$\mu$m                          & $16.54 \pm0.22$     & mJy           \\
ALMA 1250$\mu$m                         & $5.94\pm0.23$       & mJy           \\
ALMA 3300$\mu$m                         & $237\pm3$    & $\mu$Jy           \\
VLA 3GHz                        & $127.8\pm5.5$    & $\mu$Jy       \\
VLA 1.4GHz                        & $261\pm19$    & $\mu$Jy       \\
\hline
\hline
\end{tabular}\\
}
{
}
\end{table}

\subsection{ALMA dust continuum observations} 

We recovered publicly available ALMA band 7 data covering RO-1001, which consist of three pointings from projects 2015.1.00137.S (PI: N. Scoville) and 2016.1.00478.S (PI: O. Miettinen). They are imaged with a common restoring beam of 0.15 arcsec with natural weighting (given the maximum baseline of 1107\,m), then corrected for primary beam attenuation and combined. The continuum rms reaches about 50\,$\mu$Jy\,beam$^{-1}$ in the central region at the restored resolution (and about $\times$2 higher if tapered to a beam of 0.6$''$).
Three galaxies are very clearly detected (Fig.~3; Table~1). No other significant detection is present in the ALMA imaging.

In order to measure the size of the dust emission, we modeled the ALMA observations in the {\em uv} space, combining all datasets. We fitted circular Gaussian sources for simplicity: sizes for individual objects are reported in Table~1.  

\subsection{NOEMA CO observations: constraining the redshift of ALMA-detected galaxies.}

We observed the RO-1001 field with the IRAM NOEMA interferometer covering the CO[3-2] line emission redshifted to 88.3~GHz for $z=2.91$, with the main aims of confirming the redshift of cluster galaxies and measuring accurately their systemic velocities. The field was observed during November~2018 to March~2019. A total of seven tracks were obtained. The data were calibrated in a standard way using GILDAS {\it clic} software packages, and analyzed with {\it mapping}. The data has an rms sensitivity of $5\,\mu$Jy/beam in the continuum and of 24~mJy~km~s$^{-1}$ over 300~km~s$^{-1}$ for emission lines. The primary beam is about 1~arcmin, thus covering a large area around the nebula. The resulting synthesized beam at 88.3~GHz is rather elongated at $4.0\times1.8''$, with a position angle of 15$^\circ$. None of the ALMA detected sources are resolved at this resolution. We thus extracted their spectra by fitting PSFs in the {\em uv} space at the known spatial positions from ALMA. We simultaneously fit all galaxies in the field, i.e. the three ALMA galaxies and a  bright interloper falling by chance in the large NOEMA field, in order to avoid being affected by sidelobes. The continuum is strongly detected in all three ALMA sources in the RO-1001 structure. We searched for emission lines in the spectra by identifying channel ranges with excess positive emission and identified the strongest line in each galaxy spectrum as the one with the lowest chance significance (Jin et al. 2019; Coogan et al. 2018). Sources B and C have very strong CO[3-2] detections with $\mathrm{SNR}\sim14$ at $z=2.91$ and very broad emission lines with full width zero intensity of FWZI$\sim1000$~km~s$^{-1}$, { significantly broader than the \lya\ spectra at their positions (Fig.~4), reflecting the fact that they are  massive galaxies with fairly compact sizes. In the case of source~C the CO[3-2] emission is offset in velocity from the emission of the \lya\ nebula at its position. A weaker, additional \lya\ emission is observed at the same position at -2000~km~s$^{-1}$ offset velocity, outside of the range defined by the integrated \lya\ spectrum of the RO-1001 nebula. At the same velocity another individual \lya\ emitter is observed at the southern edge of the probed KCWI field, suggesting that it could be an unrelated galaxy on the line of sight. Alternatively, we could be detecting in \lya\ jet-like emission from the weak radio AGN possibly present in source~C (see next Sections). This would explain both \lya\ peaks in the spectrum at this position, on opposite velocity sides with respect to the CO[3-2] emission from source~C.
}
For source A the strongest feature in its spectrum is a fairly weak, 3.5$\sigma$ emission with 380~km~s$^{-1}$ of FWZI. While in itself its reality could be questionable, it turns out that this feature is offset by only 460~km~s$^{-1}$ from the average \lya\ velocity from the whole system ($z=2.9154$), when we cover CO[3-2] over about 50,000~km~s$^{-1}$ in total. The probability of a line with such SNR to be found by chance so close to the structure redshift is about 1\%. If we also take into account that the weak line falls almost exactly on top of the \lya\ velocity at the position of the galaxy{  (Fig.4 center-bottom panel; see also Fig.~10a), }we conclude that the identification of this weak feature as CO[3-2] is quite certain, and thus the redshift of the galaxy from the simultaneous detection of CO[3-2] and \lya. A summary of the properties of the three detections by ALMA and NOEMA is given in Table~1.

We note that there are large variations in the CO[3-2] flux to underlying continuum ratio (Table~1). This can be due to several reasons, including variations in the dust temperature and/or CO excitation ratio. They might also be connected to rapid SFR variations given that the dust continuum timescale is 50--100~Myr while the CO[3-2] line is sensitive to the instantaneous dense gas content. However, we note that galaxy A  has the lowest CO[3-2]-to-dust-continuum ratio, and is also the most compact: a factor of two smaller in radius than the other two (which are also extremely compact already). We speculate that the lack of CO[3-2] might be due to high optical depths as recently claimed for high-$z$, dusty galaxies (Jin et al. 2019; Cortzen et al. 2020).

The fourth-most massive galaxy in the system (object D) remains undetected in ALMA and NOEMA continuum and has no CO[3-2] detection. Assuming the average SED temperature as seen in the group to convert the ALMA upper limits into SFR, we place an  upper limit of $\mathrm{SFR}<30\,M_\odot$~yr$^{-1}$ (sSFR$<0.3$~Gyr$^{-1}$), which locates this galaxy one dex below the main sequence. It is thus a candidate quiescent system in the group. 

 \begin{figure}
{\centering 
 \includegraphics[width=0.49\textwidth]{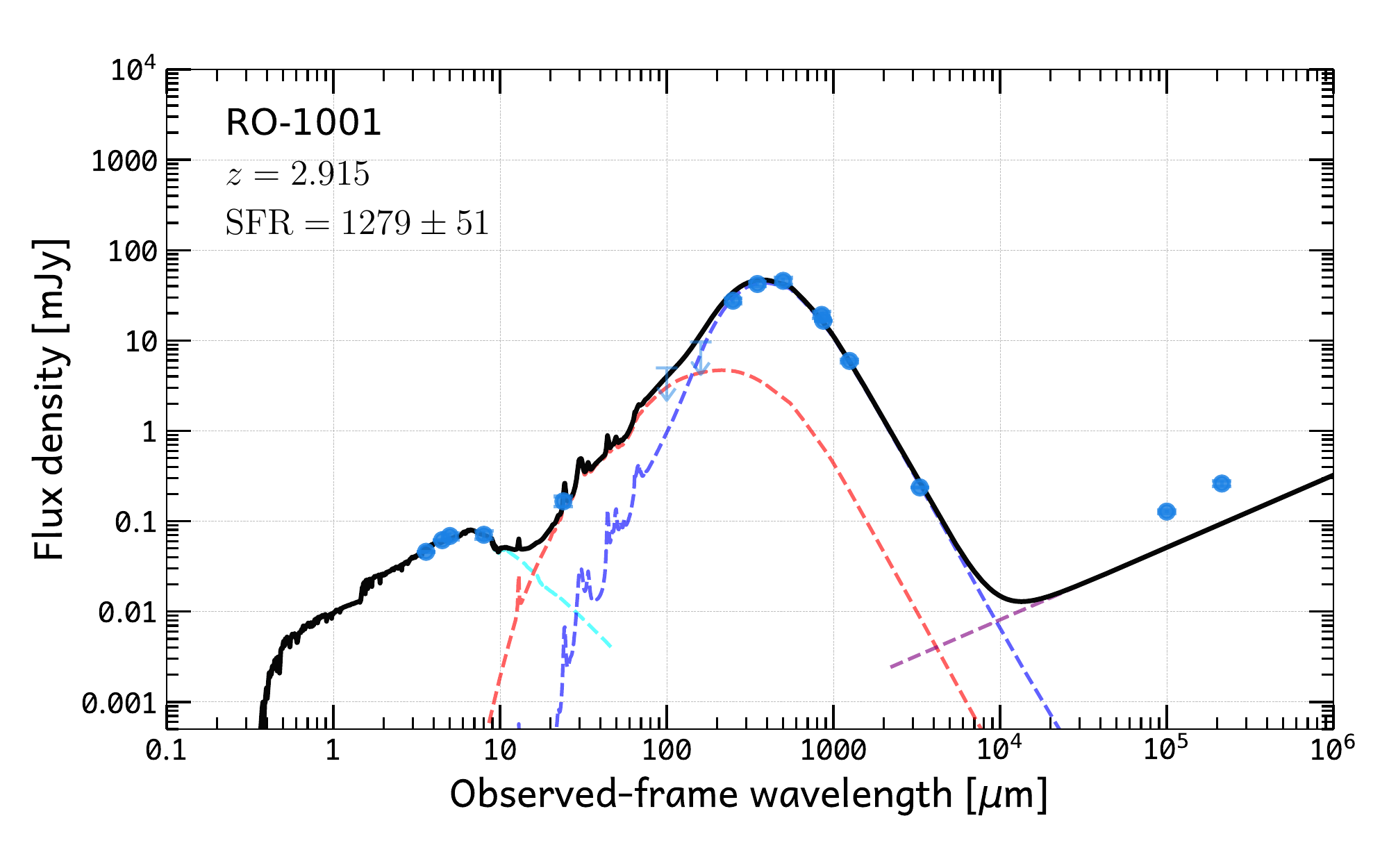}}
\caption{Integrated spectral energy distribution over the RO-1001 field { (see Table~2 for the photometric points)}. The cyan, purple and red lines are BC03 stellar template, DL07 cold (ambient) dust and DL07 warm (PDR) dust, respectively. The radio excess is within a factor of two of the average radio-infrared correlation, and is therefore not very significant, but may be associated with  galaxy 'C' that shows a somewhat elongated radio morphology suggestive of a weak jet. Instead, the SED leaves no room for the presence of AGN torus emission in the mid-IR, { with an upper limit on its bolometric luminosity of $L_{\rm AGN}<2\times10^{45}$~erg~s$^{-1}$.}
 }
\end{figure}

 \begin{figure}
{\centering 
 \includegraphics[width=0.49\textwidth]{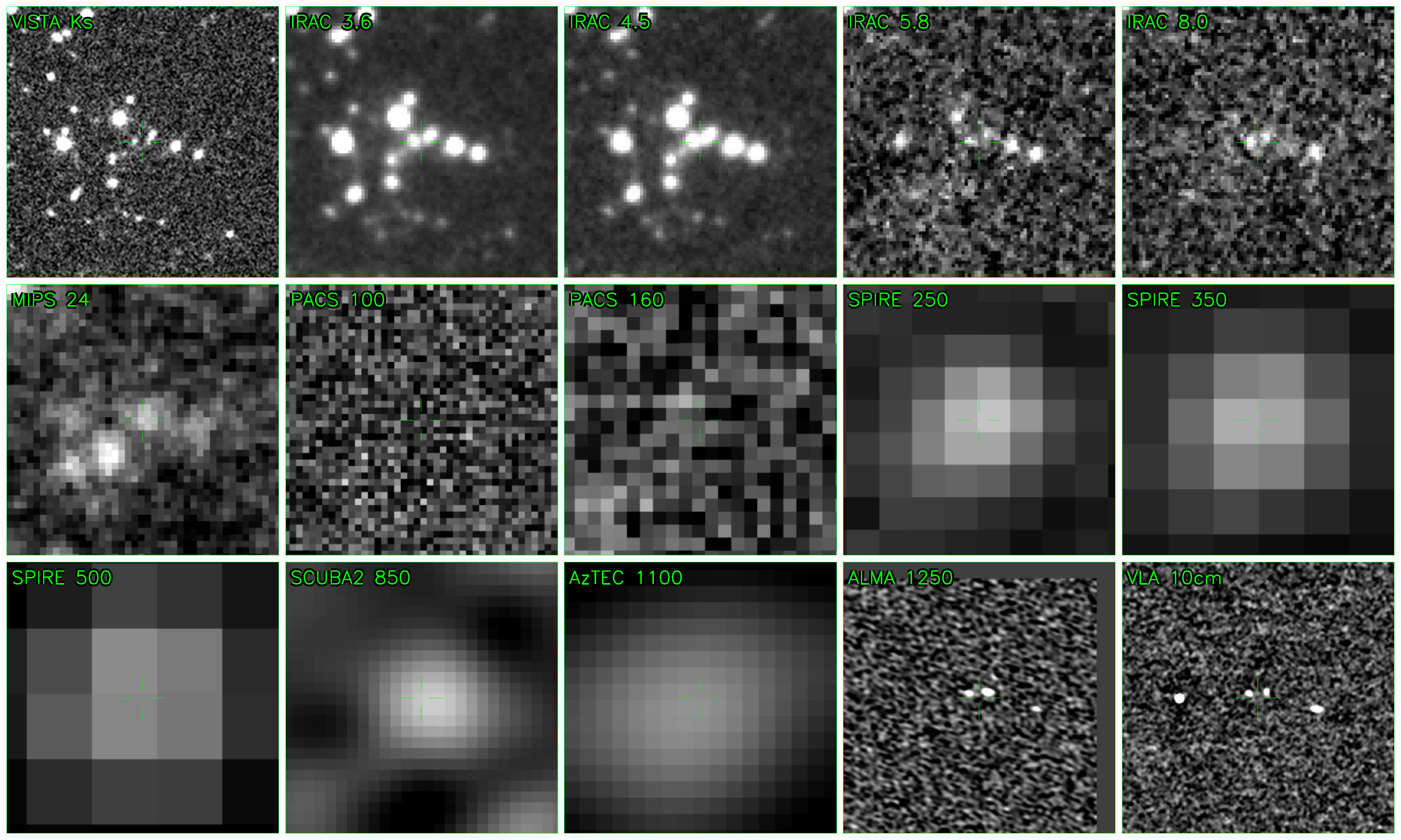}}
\caption{Multiband imaging of the RO-1001 field from Ks, Spitzer IRAC, Herschel PACS and SPIRE, SCUBA2, AzTEC, ALMA and radio (as labeled). Each cutout is $50''$ wide.
 }
\end{figure}

 \begin{figure}
{\centering  \includegraphics[width=9cm]{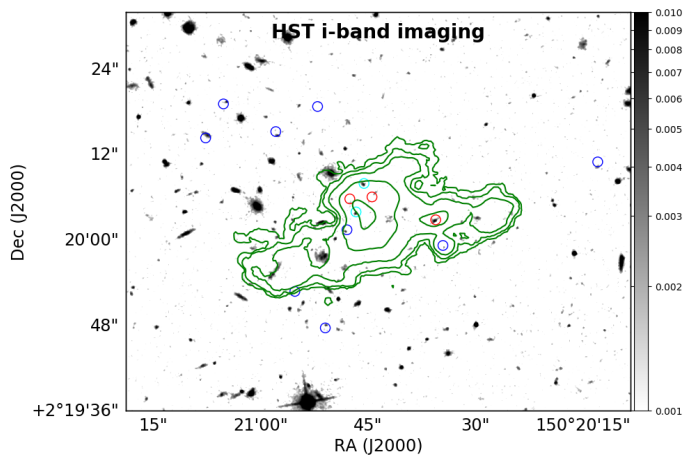}}
\caption{HST imaging of the field in the F814W filter, single orbit (Scoville et al. 2007). The image is smoothed with a Gaussian with the same FWHM of the PSF (0.1$''$) to enhance
visibility of faint features; the colorbar shows relative pixel fluxes. Green contours show the \lya\ emission. Blue circles are \lya\ emitters identified inside the nebula from the KCWI data (within 2000~km~s$^{-1}$): the small offsets with respect to the HST positions are a combined effect of the noise in the \lya\ cube affecting their recovery and the accuracy of the astrometric solution of the KCWI cube.
{ Cyan circles mark two galaxies in the structure identified from UV absorption, }red circles show the positions of the ALMA galaxies detections.
 }
\end{figure}

\subsection{Integrated star formation activity.}

The total IR luminosity of the group is derived by fitting (Jin et al. 2018) Herschel, SCUBA2, and ALMA and NOEMA continuum fluxes. The Herschel/SPIRE images are fitted using two PSF components on the images on the position of sources C and the average position of A and B, respectively. Due to the higher spatial resolution, we fit  at the position of all three ALMA sources in the SCUBA2 image (Figs.~5; 6). 
We fit the SED of the group, including the summed photometry from Spitzer, Herschel, SCUBA2, ALMA and NOEMA, and obtain a best fitting SFR~$=1200\ {\rm M}_\odot$~yr$^{-1}$ for a SED with an average intensity of the radiation field $\langle U\rangle = 45$ (Fig.~5). Both the dust temperature and average specific star formation rate (sSFR; 3~Gyr$^{-1}$ on average over the three ALMA galaxies) are in agreement with those of typical main sequence galaxies at $z\sim3$ (Bethermin et al. 2015; Schreiber et al. 2018). Assuming that individual SFRs of the ALMA-detected galaxies scale like the ALMA/NOEMA continuum fluxes, we conclude that also the  sSFRs of the individual galaxies place them within the main sequence.  This does not imply that they are normal galaxies, as the very compact ALMA sizes betray some ongoing/past starbursting activity (Puglisi et al. 2019).

The ALMA 870~$\mu$m emission from the three detected galaxies is consistent within the uncertainties with the SCUBA2 signal  (Fig.~5), implying that the bulk of the IR emission and SFR in the RO-1001 group comes from the three ALMA detections.

We identified additional star-forming galaxies in the structure at $z=2.91$  via their \lya\ emission, or in two cases via possible UV absorption lines (Fig.~7). They are not detected in the near-IR to current depths, impliying they are lower-mass, star-forming galaxies. { Their contribution to the integrated SFR in the RO-1001 group, derived from the \lya\ emission, is negligible. }

\subsection{AGN limits.}

A cross-match between RO 1001 and the deepest Chandra COSMOS+Legacy images (Civano et al. 2016)  yields no X-ray point-source detection. We stacked the observed soft (0.5--2 keV) and hard (2--10 keV) bands at the position of the three ALMA sources using CSTACK\footnote{CSTACK is publicly available at: \url{http://lambic.astrosen.unam.mx/cstack/}}. We estimate an average $L_{\rm X}<3.5\times10^{43}$~erg~s$^{-1}$ (3$\sigma$ upper limit) in the   rest-frame 2--10~keV directly from the  soft X-ray fluxes (0.5--2~keV observed). The observed hard X-ray emission maps directly into 8--40~keV rest-frame and is much less affected by obscuration. It provides a limit of $L_{\rm X}<5.4\times10^{43}$~erg~s$^{-1}$ in the  rest-frame 2--10~keV, K-corrected assuming a power-law X-ray spectrum with photon index $\Gamma=1.4$ (Gilli, Comastri \& Hasinger 2007). When spread over the three sources, this gives an integrated limit of  $L_{\rm X}<1.5\times10^{44}$~erg~s$^{-1}$ in the  rest-frame 2--10~keV.
We further estimated 3$\sigma$  $L_{\rm X}$ upper limits of $L_{\rm X}<7\times10^{42}$~erg~s$^{-1}$ (rest-frame 0.5--2~keV) corresponding to an AGN bolometric luminosity of $L_{\rm AGN}<2\times10^{45}$~erg~s$^{-1}$, fairly independent of obscuration.
 This is corroborated by analysis of the individual broad-band SEDs (Jin et al. 2018), in which the mid-IR AGN component is always negligible relative to the host galaxy, providing similar limits on any possible AGN bolometric luminosity. 

{ We  calculated the typical AGN luminosity expected from the integrated SFR of the group/cluster, assuming empirical $M_{\star}$-dependent BHAR/SFR relations (Mullaney et al. 2012; Rodighiero et al. 2015; Delvecchio et al. 2020) for star-forming galaxies. At the high mass end, the average black hole accretion rates scales as BHAR$\sim6\times10^{-4}\times SFR$ (see Fig.4 in Delvecchio et al. 2020). Estimating in this way the BHAR using the integrated SFR in the group, and converting it into a bolometric AGN luminosity, we would expect from this structure an average AGN activity at the level of  L$_{\rm AGN}\sim$2$\times$10$^{45}$~erg~s$^{-1}$, comparable to the upper limits inferred from the direct estimates in the X-rays. We use this bolometric luminosity for estimates of energetics.}

Using  modeling  that reproduces the evolution of the X-ray luminosity function through cosmic time on the basis of the mass function and SFR distributions statistically observed in galaxies (Delvecchio et al. 2020), we infer that, given the massive galaxies in the RO-1001 structure and their SFRs, the probability of observing one of them with QSO luminosities as high as those in Borisova et al. 2016 is $\sim 10^{-4}$.  This argues that there was only a very brief interval, at best, during which our RO-1001 group might have plausibly been selected in one of the QSO nebulae surveys.

The RO-1001 structure was selected due to the presence of   three VLA-detected sources at 3~GHz. All have moderate radio power $L_{\rm 3\,GHz} \sim10^{24.2-24.6}$~W~Hz$^{-1}$. Given the integrated SFR of the group/cluster ($\sim1200\,M_\odot$~yr$^{-1}$) and a redshift-dependent infrared-radio correlation, radio emission can be broadly explained by consistent levels of star formation
within less than a factor of 2 (Fig.~5). { Careful inspection of the shape of image of Source C (see insert in Fig.8) shows that this source appears to be elongated along the E-W direction in the radio at 3~GHz. This suggests that some extra emission (e.g., from  jets) might be present, which might trace (past or relatively weak) AGN activity in this source. This is consistent with the weak \lya\ (Fig.~1,~2) and HST i-band continuum (Fig.~7) detections of this galaxy and, potentially, with the double peak spectrum in Fig.4.}

\begin{figure}
{\centering  
\includegraphics[width=0.48\textwidth]{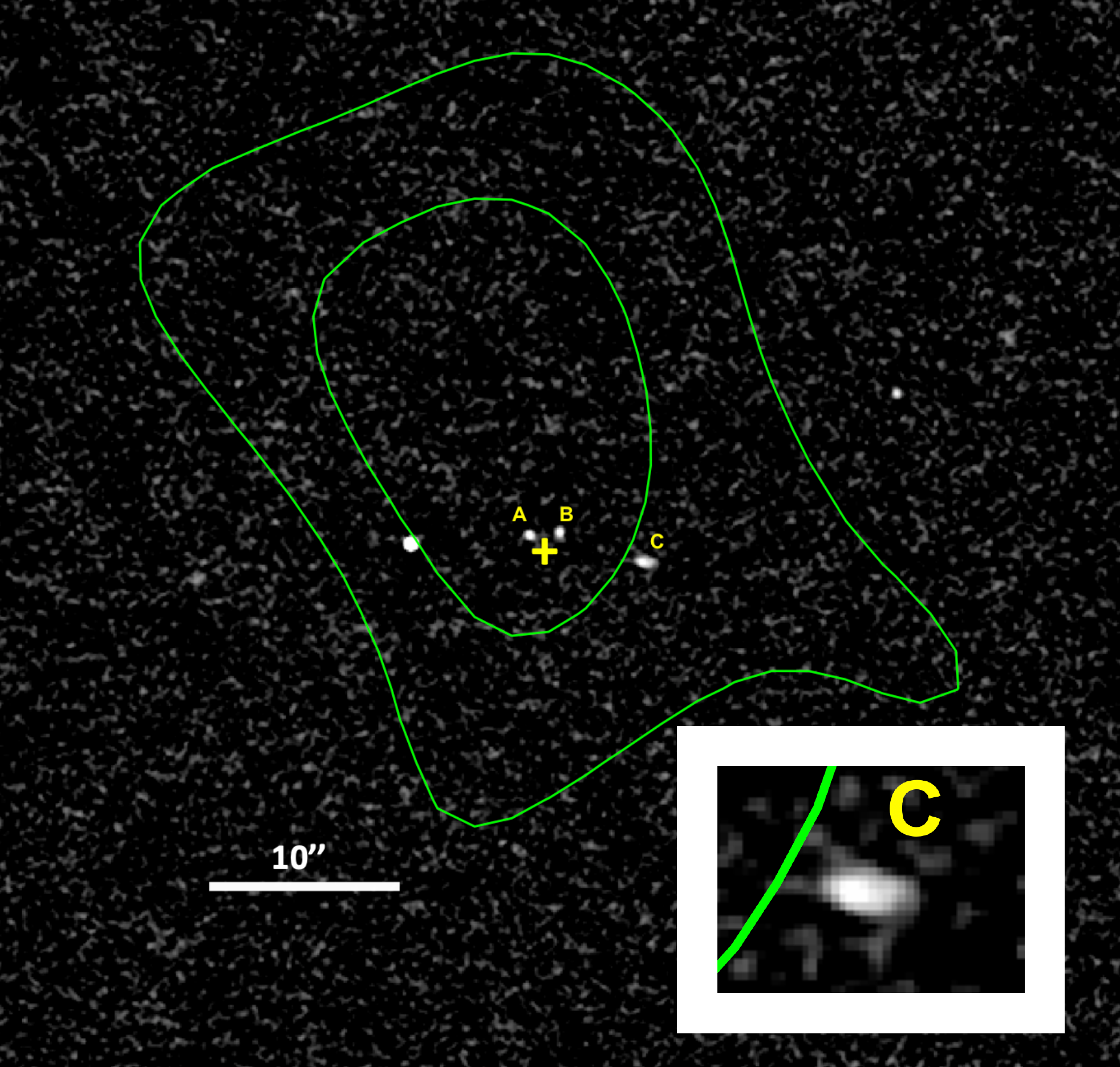}
}
\caption{
X-ray contours (1 and 2$\sigma$) from Chandra+XMM overlayed over the VLA 3GHz image of the RO-1001 field. The offset between the X-ray peak and the \lya\ peak (yellow cross) is entirely consistent to arise from noise. The 3.1$\sigma$ signal measurement is obtained integrating the emission over a region centered at the RO-1001 centroid and with 24$''$ radius (set to avoid a bright, unrelated point X-ray source that is increasing the noise substantially). { The insert in the bottom-right corner shows a zoom of source~C, allowing to appreciate its elongated shape. }The bright, left-most radio source is a low-redshift interloper.
}
\end{figure}

\subsection{X-ray constraint and halo masses}

In order to constrain further the hosting halo mass, we searched for extended X-ray { bremsstrahlung}  emission from the hot gas at the position of RO-1001 (lacking any detectable point source X-ray emission, see previous section).
We used an X-ray image in the 0.5--2~keV range, produced by combining the 
Chandra and XMM-Newton images after background and point source removal (Fig.8). 
We used a 24$''$ radius aperture to place the flux estimates on the 
source, obtaining a value of $5.8\times10^{-16}$ erg s$^{-1}$ cm$^{-2}$,
which is a 3.1$\sigma$ excess over the background. The source has a 
230~ks  Chandra exposure, with a corresponding upper limit on the point 
source contamination of $1\times10^{-16}$ erg s$^{-1}$ cm$^{-2}$ (using 
the same count-rate to flux-conversion-rate as for the source).
At a redshift of $z=2.9$ this corresponds (Leauthaud et al. 2010) to rest-frame 0.1--2.4 keV $L_\mathrm{X}$ of 
$1.1\times10^{44}$ erg s$^{-1}$ and a mass of $M_{\rm 
DM}\sim4\times10^{13}M_\odot$. The temperature of the ICM implied by the same correlations is $\sim2$~keV.

 \begin{figure}
{\centering  \includegraphics[width=7.2cm,angle=-90]{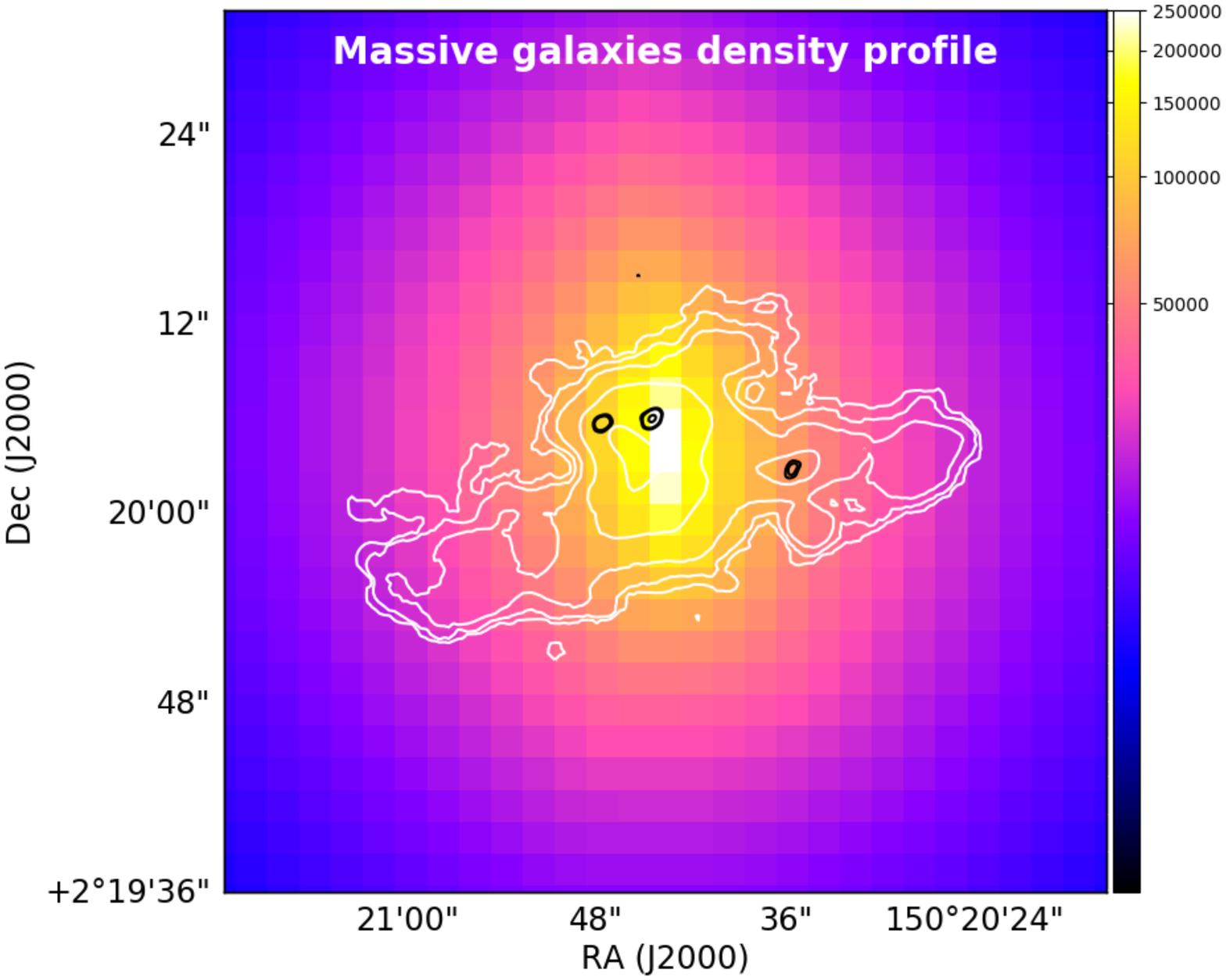}}
\caption{
The 4$^\mathrm{th}$-nearest neighbor projected density map of massive galaxies in RO-1001. Black contours are from ALMA detections. Notice the excellent coincidence of the \lya\ emission (contours) with the potential well of the RO-1001 halo as traced by its massive galaxies. The color scale is in units of number of galaxies per square degree.
}
\end{figure}

\section{Results: general and spectral properties of the \lya\ emission nebula in RO-1001 and comparison to cold accretion predictions}

The RO-1001 structure appears to be a group hosted in a single, fairly massive dark matter halo. The halo mass of RO-1001 is estimated using three methods returning consistent results: 
\begin{enumerate}
\item   The integrated stellar mass of 5.4$\times10^{11}M_\odot$ of its four most massive galaxies (Fig.~2; Section 2.2) above a mass completeness limit of ${\rm log(M}/M_\odot)>10.8$  
and scaled based on stellar to DM ratios corresponds to $M_{\rm DM}\sim$4--6$\times10^{13}M_\odot$.
\item
Herschel+ALMA reveal a total infrared luminosity $L_{\rm IR}=1.2\times10^{13}L_\odot$ from star formation (Sect.2.5): scaling from the $z=2.5$ cluster CL-1001 (Wang et al. 2016) this   returns $M_{\rm DM}\sim4\times10^{13}M_\odot$ (accounting for the expected cosmic increase of the SFR over $z=2.5$--2.9).  
\item A blind (no free parameters) X-ray measurement centered at the barycenter of the 4 massive galaxies returns a 3.1$\sigma$ excess over the background corresponding to $M_{\rm DM}\sim4\times10^{13}M_\odot$  (Sect.2.7;  contamination by X-rays from star formation is negligible, while no individual point-like X-ray sources are present). 
\end{enumerate}
RO-1001 thus appears roughly half as massive to Cl-1001 at $z=2.51$ (Wang et al. 2016), the previously most distant X-ray structure known,
but more distant,  at $z=2.91$. This is 400~Myr earlier (15\% of Hubble time), which is substantial because at these epochs the halo mass function is rapidly evolving (see e.g., Mo \& White 2002).

This mass estimate is important  to quantitatively compare our observations and overall properties of its \lya\ nebula (Sect.~2.1) to predictions from cold accretion theory. In fact, 
RO-1001 falls close to the $z$-$M_{\rm DM}$ regime where cosmological cold flows might penetrate the hot halo, as inferred from simulations and models (Dekel et al. 2009; Behroozi et al. 2018; cold gas would not penetrate at lower redshifts). Substantial accretion of cold gas in this structure would not be surprising, given the ongoing  $\mathrm{SFR}=1200\,M_\odot\,\mathrm{yr}^{-1}$ of the group galaxies. 
Theory predicts  gas accretion from the IGM scaling as $M_{\rm DM}^{1.15}\times(1+z)^{2.25-2.5}$  (Dekel et al. 2013; Neinstein \& Dekel 2008), or about $10,000\,M_\odot$~yr$^{-1}$ in RO-1001. This would be  sufficient to feed the ongoing SFR, provided that a non-negligible fraction of the likely multi-phase inflow (Cornuault et al. 2018) remains cold. Additionally, local cooling due to the interactions and shocks between the infalling gas and the hot cluster gas might also provide  the required cold gas fuel for star formation (Mandelker et al. 2019; Zinger et al. 2018). 

In the light of this, we present in the following  an investigation of \lya\ emission observables in RO-1001, with a comparison to expectations of cold accretion models informed by the DM halo constraints, but also discussing whether the observations could be explained in an alternative outflow scenario. 

\subsection{\lya\ geometry}

The \lya\ emission does not appear to be accurately centered on individual galaxies (Fig.~2): the luminous core peaks in an empty region located at the center of the halo potential well (defined as the barycenter of the stellar mass distribution; Fig.~9), { about $2''$ (15kpc) away from galaxies A and B. This is similar to the median offset observed in local clusters between the X-ray centers and brightest cluster galaxies (BCG; Lauer et al. 2014;  15\% of the BCGs are offset by more than 100~kpc). However, it is worth noticing that RO-1001 does not contain a dominant, massive galaxy in which we could obviously recognize a BCG: the 4 massive galaxies that we have identified  have all quite comparable stellar masses within less than a factor of two. 
This is similar to what was found previously for well studied high redshift clusters  like CL-J1449, where some evidence exists though that a proto-BCG might be in the process of assembling at $z=2$ (Strazzullo et al. 2016), and even more in CL-J1001, where a dozen similarly massive galaxies are packed in a 100~kpc core at $z=2.51$ (Wang et al. 2016). This highlights a shortcoming of current cluster/group formation simulations, where BCGs are  assembled much earlier (e.g. Saro et al. 2009; Tremmel et al. 2019). This is evident also in the Rosdahl \& Blaizot (2012) simulations, where their most massive $10^{13}M_\odot$  halo already displays a prominent BCG at $z=3$. }

{ The core of the nebula, where the filaments are less collimated and merge losing their identity, and where the emission is strongest, extends over a radius of about 40-50~kpc. This corresponds to 15-20\% of the virial radius of its hosting DM halo ($R_{\rm V}\sim280$~kpc). This is quantitatively consistent with the behavior predicted for cold streams (Danovich et al. 2015) where a 'messy' interface region is expected to gradually form within 0.3$R_{\rm V}$.}
The 150--200 kpc radius traced by filaments corresponds to 50--70\% of the virial radius of its hosting DM halo, not including possible projection effects.
{ The  disappearance of the filaments beyond the virial radius is predicted by cold accretion models, as due to the the lack of hot gas compressing the cold material outside of $R_{\rm V}$ (Dekel et al. 2009; Dijkstra \& Loeb 2009; Rosdahl \& Blaizot 2012).} The filaments in RO-1001 have  substantial transverse diameters (50--70 kpc; $\sim20$\% of the virial radius). This is predicted by cold accretion models for flows into the most massive halos, broadened due to their initially higher pressure and instabilities (Cornuault et al. 2018; Rosdahl \& Blaizot 2012). 
The average surface brightness in the filaments is of order  $1\times10^{-18}$~erg~s$^{-1}$~cm$^{-2}$~arcsec$^{-2}$, with a total area above this surface brightness limit of  210~arcsec$^2$ ($1.3\times10^4$~kpc$^2$), quite comparable to theoretical expectations from cold accretion models given the hosting halo mass (e.g., see Fig.11 of Rosdahl \& Blaizot 2012). The circularly averaged surface brightness profile follows $r^{-2.1}$ at large distances, consistent with the same predictions.

{ It is interesting to consider whether the \lya\ nebula geometry, in the hypothesis of the lack of substantial cold accretion, could instead be produced by the outflow activity, and subsequent gas returning to the halo center in the form of galactic fountains (Tumlinson et al. 2017). In this scenario, the filaments would have to be interpreted as extreme AGN-driven outflows originating from (possibly different) individual galaxies (star-formation driven outflows would presumably trace multiple directions from each individual galaxy). The core could instead be interpreted as a region of filaments overlap as well as the place towards where gas ejected with velocities lower than the escape one would fall back.}
 
\subsection{\lya\ moment maps}

We can search for further diagnostics of the ongoing physical processes from the \lya\ spectral properties. The moment~1 (velocity) map of the \lya\ emission in RO-1001 is shown in Fig.~10a. Zero velocity corresponds to the average, flux-weighted redshift of the \lya\ emission ($z=2.9154$), which as a result coincides with the velocity of most of the core of the nebula, where \lya\ intensity is higher. 
The  velocity relative to the core increases in absolute value towards the outer region of the filaments reaching 400--500~km~s$^{-1}$ (and possibly up to 600--700~km~s$^{-1}$ when statistically correcting for the unknown inclinations), similar to the virial velocity and as predicted by theory within cold accretion models (Rosdahl \& Blaizot 2012; Cornuault et al. 2018), given the estimated DM halo mass. 
In the case of  infall along the filaments, our observations  would suggest that the initial (virial) velocity is progressively reduced as the flows proceed into the hot medium with which they interact, or betray the changing direction of the gas filaments while inspiraling towards the center of the potential well. 
Indeed,  inflow models (Danovich et al. 2015; Mandelker et al. 2020a) do not predict a reduction in the absolute speed of the filaments while moving from the outskirts to the core but do predict projection effects where the gradient arises from the bending of the filaments inside the halo.  This seems more consistent also with the kinematics of 
the SE filament that shows evidence for a local velocity gradient from the velocity map (Fig.~10a).  If due to rotation, this might also  represents a lower-mass dark matter halo in the process of merging into the larger system. Intriguingly, we have currently no obvious individual galaxy associated with this putative sub-halo.

The moment-2 (velocity dispersion; Fig.~10b) map shows typical local velocity dispersions in the range of 200--300~km~s$^{-1}$ (de-convolved by the instrumental resolution), 
about half of the virial velocity (classifying RO-1001 as a dynamically {\em cold} nebula) but higher than the expected thermal broadening of the cold $\sim10^4$~K gas (few tens km~s$^{-1}$),  as expected  for multiphased, cloudy accretion flows in which streams do not remain highly collimated (McCourt et al. 2018; Cornuault et al. 2018).
The observed velocity gradient between the edge of the filaments  and the core together with the velocity dispersion increasing towards the core allows us to obtain a rough estimate of the cold gas mass flow, in the scenario that we are indeed observing cold accretion. Assuming that the initial kinetic energy of the cold gas ($\dot{M}v_{\rm virial}^2$)  is partly transformed into turbulence ($\dot{M}\sigma_{\rm turbulence}^2$) and partly radiated away (mainly via \lya). Developing such calculations suggest that 10--30\% of the initial energy is converted into turbulence (Cornuault et al. 2018), with a cold mass flow rate of 1000--2000 $M_\odot$~yr$^{-1}$, thus a penetration efficiency (as defined by Dekel et al. 2013) of order of 10--20\%. This infall rate  approximates the ongoing SFR, but only a fraction of this would  reach the galaxies, consistent with the requirement that the system must be heading to a downfall of the activity in a few dynamical timescales (a few Gyr), and eventually quench (e.g., by $z<2$).

{ 
Quantitatively comparing these kinematic features to expectations from an outflow scenario is hampered by the fact that we are not aware of detailed modelling of \lya\ emission filaments produced by outflows from AGN or star formation in galaxies. However, qualitatively it appears that any attempt to reproduce these feature, and in particular the  filaments, with AGN outflows would require substantial fine-tuning. Those AGN outflows would be expected to be observed with of order 1000-2000~km~s$^{-1}$  velocities (e.g., Kakkad et al. 2020), and there should be at least three prominent ones currently prominently visible with presumably different launching directions. All of this seems hard to reconcile with the moderate velocity dispersion of the \lya\ nebula, whose full velocity range at zero intensity hardly reaches 1000~km~s$^{-1}$. In terms of velocities, the velocity gradient observed from the core to the outer edges of the two most prominent filaments, at the level of 400-500~km~s$^{-1}$ would also be complex to explain. 
 }

\begin{figure*}
{\centering  
\includegraphics[width=0.99\textwidth]{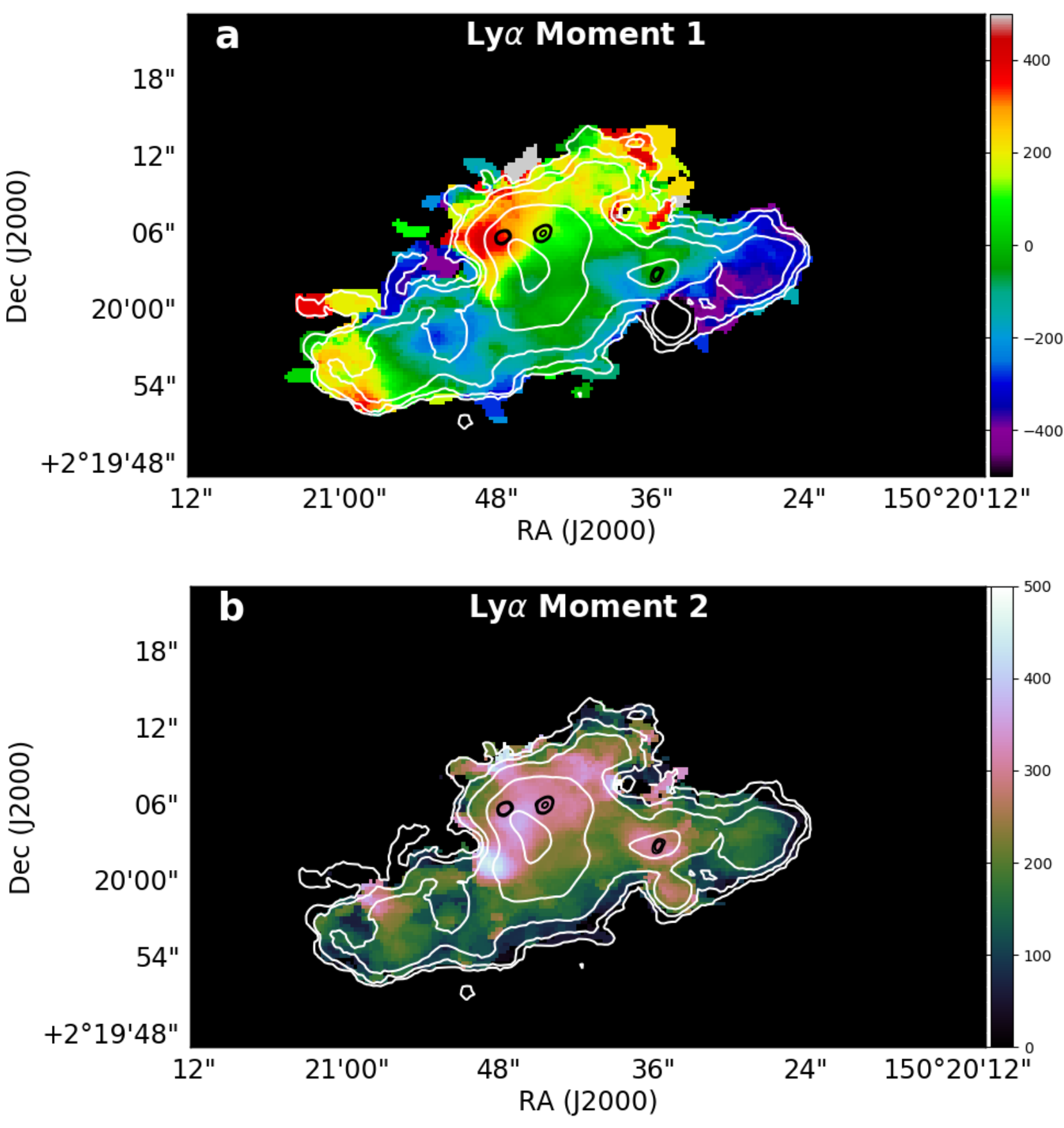}
}
\caption{
\lya\ velocity  map (top; moment~1) and velocity  dispersion map (bottom; moment~2; de-convolved by instrumental resolution). 
Color bar levels of both maps (right scale) are expressed in km~s$^{-1}$. We only show pixels where moment errors are below 50~km~s$^{-1}$. 
ALMA continuum sources  are shown as black contours, in both panels.
White contours show the \lya\ surface brightness levels from Fig.~2. 
}
\end{figure*}

\begin{figure*}
{\centering  
\includegraphics[width=0.89\textwidth]{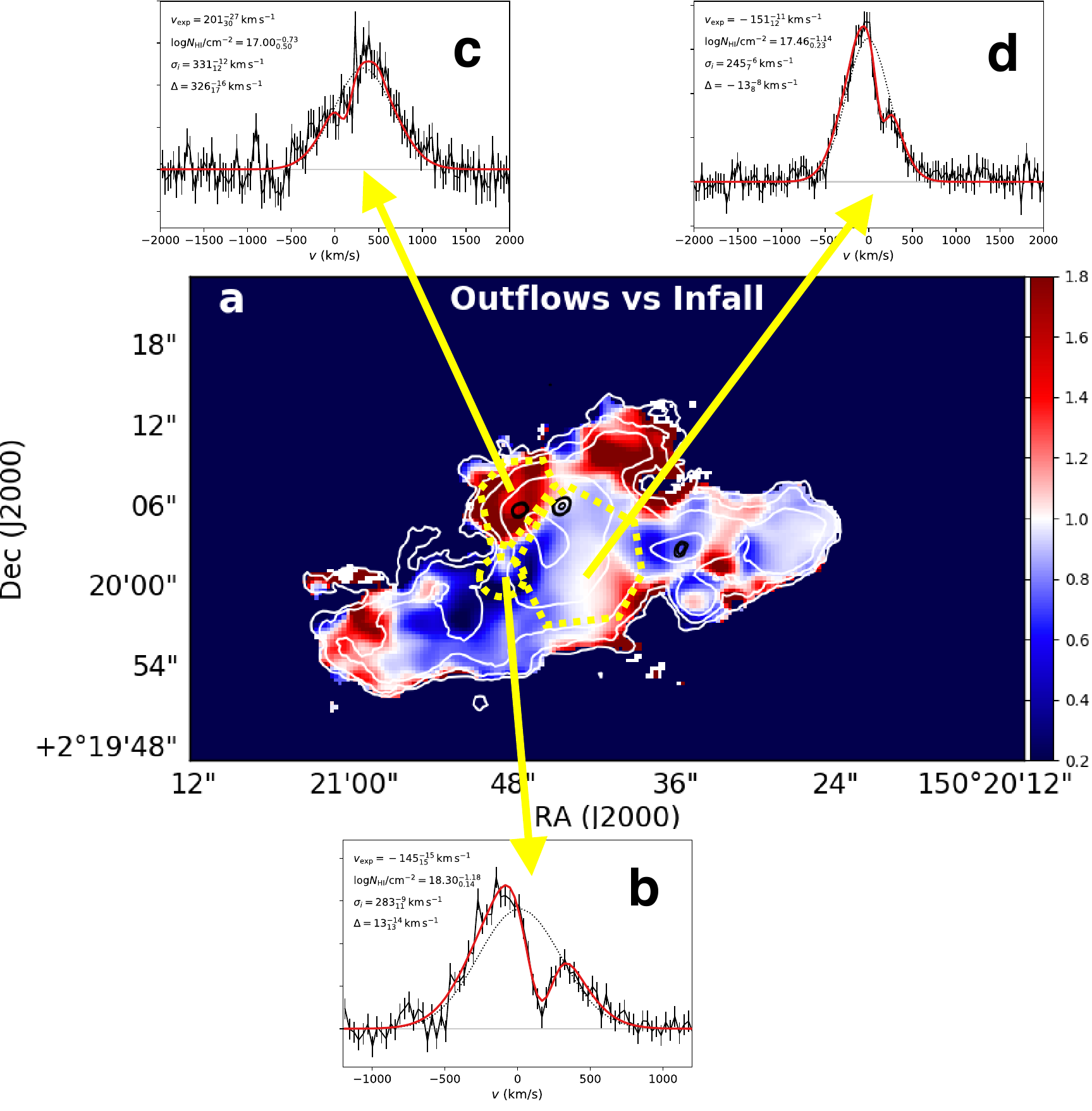}
}
\caption{
The color coding in the image (right scale) shows the ratio between redshifted and blueshifted components integrated flux (Panel~a) from a ROHSA analysis of the deeper low  spectral resolution data. Blue (red) colors correspond to infall- (outflow-) dominated regions.
ALMA continuum sources  are shown as black contours.
White contours show the \lya\ surface brightness levels from Fig.~2.  Yellow-dotted contours show the regions where spectra were extracted, using the higher spectral resolution data, as shown in Panels~b,c,d. {  Observed spectra are shown as black continuous lines, while best fitting modeling from the shell models in Sect.3.3 are shown in red for the observed spectrum and dotted black for the intrinsic spectrum}. Resulting constraints on physical quantities are labeled 
($v_{\rm exp}$ is the bulk velocity of the shell; $\Delta$ is the systemic velocity). The spectrum in Panel~d shows a prominent blue peak and is integrated over most of the core of the nebula, with the exception of the NE corner whose spectrum is shown in Panel~c and is characterized by a prominent red peak. Panel~b shows the region with the largest velocity dispersion from Fig.~10b. 
}
\end{figure*}

\subsection{\lya\ radiative transfer modeling.}

The typical spectra of the \lya\ emission in the core of the nebula is reported in Fig.~11d, showing a double-peaked shape,  with the blue component stronger than the red one. 
This is reversed from what is observed in most \lya\ emitting galaxies at both low- (Kunth et al. 1998; Henry et al. 2015; Yang et al. 2016; Rivera-Thorsen et al. 2015) and high-redshift (Erb et al. 2014; Orlitov{\'a} et al. 2016; Matthee et al. 2017; Herenz et al. 2017) where the observation of a stronger red peak is associated with outflows (Kulas et al. 2012), hence here suggesting infall (see Ao et al. 2020 for a \lya\ blob showing similar spectral properties).

{ We modeled the spectra accounting for radiative transfer effects, that is, we take scattering, frequency
redistribution, as well as scattering out and back into the
line-of-sight into account.
To do so, we employ radiative transfer models computed using the
Monte-Carlo radiative transfer code {\it tlac} (Gronke \& Dijkstra 2014)
which follows the trajectories of individual
`photon packages' through frequency- and real-space. We assume the
hydrogen and dust to be be located in a shell surrounding the emission
site.
This `shell model'  (Ahn et al. 2003; Verhamme et al. 2006) consists of a moving concentric shell of neutral
hydrogen  and dust which surrounds a central \lya- (and continuum-)
emitting source. The appropriate HI absorption cross sections (depending on relative velocity and also temperature) are implemented.} This setup introduces at least five parameters: the
neutral hydrogen column density $N_{\mathrm{HI}}$, the dust optical
depth $\tau_{\mathrm{d}}$, the velocity of the shell
$v_{\mathrm{exp}}$ (defined to be positive when outflowing), the
(effective) temperature of the shell $T$ (incorporating any
small-scale turbulence), and the width of the intrinsic Gaussian line
$\sigma_{\mathrm{i}}$. This is admittedly simplistic but providing qualitatively similar results to spherical clouds, with either an illuminating central point source or uniformly distributed in the volume  (Verhamme et al. 2006), and even a multiphase medium with high clumping factor as discussed above (Gronke et al. 2017). 
Specifically, we fitted the observed (continuum-subtracted) \lya\
spectrum using an improved pipeline originally described in Gronke et al. (2015). This pipeline consists of $12,960$ models
covering the $(N_{\mathrm{HI}},\,v_{\mathrm{exp}},\,T)$ parameter
space, and the other parameters are modeled in post-processing. To the
five parameters above, we also added the systemic redshift which we
allowed to vary within $[-700,\,400]\,\mathrm{km}\,\mathrm{s}^{-1}$  of $z=2.9154$ (corresponding to the velocity range spanned by individual CO[3-2] detections in the group). Furthermore, we modelled the effect of the spectral
resolution by smoothing the synthetic spectra with a Gaussian with
FWHM of $150\,\mathrm{km}\,\mathrm{s}^{-1}$, equal to the resolution of the BM dataset.
In order to sample the posterior distribution sufficiently, we used
the tempered affine invariant Monte Carlo sampler of the \texttt{Python} package \texttt{emcee} (Goodman \& Weare 2010)
We chose to employ $10$ temperatures, $200$ walkers, and $2000$ steps that we found to be sufficient to sample the posterior.

For all fitted spectra (panels b, c, d in Fig.~11) the fitting returns moderately low column densities of neutral hydrogen, at the level of  $10^{17-18.5}$~cm$^{-2}$, { suggesting fairly reduced radiative transfer effects that cause the observed \lya\ profile to be weakly distorted  respect to the intrinsic profiles (dotted black lines in panels b, c, d in Fig.~11). This would support the conclusion that the inferred systemic velocities and dispersions previously measured from the \lya\  moments are reliable (meaning that moments computed on the observed  and on the intrinsic \lya\ profiles are very close), as often observed in \lya\ blobs (Herenz et al. 2020).} 

The blueshift-dominated spectral shape requires an overall average infalling velocity of $v_{\rm infall}\sim150$~km~s$^{-1}$. 
The systemic velocity  is close to our zero-velocity scale, which is defined as the \lya\ flux weighted velocity of the full nebula, and is  consistent with the average CO[3-2] redshifts of the three ALMA galaxies. { This is consistent with  such emission originating from gas at rest with respect to the center of the potential well, on average.} Observations of non-resonant lines would be decisive to confirm this feature and all results from shell modeling, in general.

Several hotter regions with 400--500 km~s$^{-1}$ dispersion are evident (red spots in Fig.~10b): it is tempting to interpret them as possible shock fronts from the incoming material surrounding the core regions where gas density is highest. Modeling the spectra supports this idea, recovering higher HI column densities at one of the most conspicuous locations (Fig.~11b).

For a region around the galaxy `A' the spectral shape is reversed (Fig.~11c), with the red component being stronger. This red-dominated spectrum is fitted as originating at a systemic velocity of 326~km~s$^{-1}$; shifted in the direction of  the CO[3-2] velocity inferred for the underlying galaxy `A' (460~km~s$^{-1}$). It is difficult to distinguish such a velocity offset as due to relative motion with respect to the center of the system, or due to a Hubble flow effect that would require a physical separation of about 1~Mpc along the line-of-sight.
The prevalence of the red peak suggests in any case, an overall outflowing velocity of $v_{\rm outflow}\sim200$~km~s$^{-1}$. The red-core emission is thus best understood as coming from outflows likely originating from galaxy `A'.
If this is the case, the modest velocity offsets and flow-velocities suggest star formation-driven outflows (consistent with our estimates in Table~3), given that for AGN driven ones some 1000--2000~km~s$^{-1}$ velocity offsets would be expected.

\subsection{\lya\ double Gaussian decomposition.}

It would be very instructive to be able to carry out a detailed spectral analysis also along the filaments, evaluating whether their spectral shape is also mostly blueshifted. However, they are detected at high S/N only in the lower spectral resolution data. 
 We therefore decompose pixel-by-pixel  with double Gaussians  the lower-resolution  \lya\ spectra over the whole nebula, using \texttt{ROHSA}   (Marchal et al. 2019), 
a multi-Gaussian decomposition algorithm based on a regularized nonlinear least-square criterion that takes into account the spatial coherence of the emission. Here we choose to decompose the signal into a sum of $N=2$ Gaussians, in order to capture the basic effects of resonant scattering, producing blueshifted and redshifted components (Fig.~11). 
The model of the emission is then
\begin{equation}
  \tilde I\big(\nu, \thetab(\rb)\big) = \sum_{n=1}^{N=2} G\big(\nu, \thetab_n(\rb)\big)
  \label{eq::model_gauss}
,\end{equation}
with $\thetab(\rb) = \big(\thetab_1(\rb), \dots, \thetab_n(\rb)\big)$ and where
\begin{equation}
  G\big(\nu, \thetab_n(\rb)\big) = \ab_n(\rb) \exp
  \left( - \frac{\big(\nu - \mub_n(\rb)\big)^2}{2 \sigmab_n(\rb)^2} \right)
\end{equation}
is parametrized by $\theta_n = \big(\ab_n, \mub_n, \sigmab_n\big)$ with
$\ab_{n} \geq \bm{0}$ being the amplitude, $\mub_{n}$ the position, and
$\sigmab_{n}$ the standard deviation 2D maps of the $n$-th Gaussian
profile. The estimated parameters $\hat{\theta}$ are obtained by minimizing the cost function as described in Marchal et al. 2019. The latter includes a Laplacian filtering to penalize the small-scale fluctuation of each 2D map of parameters. Note that in order to perform this minimization, the whole emission cube is fitted at once. The strength of this filtering is controlled by three hyper-parameters $\lambda_{\ab}$=10, $\lambda_{\mub}$=4000, and $\lambda_{\sigmab}$=4000. These parameters have been empirically chosen to obtain a spatially coherent solution of $\hat{\theta}$ with the smallest residual $\tilde I\big(\nu, \theta(\rb)\big)$ - $I\big(\nu)$. We find a median absolute residual of 2.3\%, showing that $N$=2 provides a good spatially coherent description of the signal. 

We obtain spatially resolved maps of the ratio of redshifted to blueshifted components (Fig.~11a), that can be used to obtain insights on where infall vs. outflows might prevail.  The result of this analysis is that infall could dominate also along the two main filaments, except for a few regions that appear to be instead dominated by outflows, similarly to the core (including possibly the whole Northern filament). { These spectral profiles suggest that the filaments are globally infalling. However, a definitive confirmation can come only with higher spatial and spectral resolution spectroscopy of the same regions. }

\begin{table*}
{
\caption{Gas flows, energetics and \lya\ production}
\label{tab:2}
\centering
\begin{tabular}{c|ccc|ccc}
\hline\hline
                & \multicolumn{3}{c}{RO-1001 nebula}                                                                                                                                                                                                     & \multicolumn{3}{c}{QSO nebulae} \\
Energy  source  & Constrain                                                                                                      & Total          & Effective                                                                                            & Constrain                               & Total    & Effective \\
\hline
AGN photoionization      & $L_{\rm AGN}\lesssim2\times10^{45}$~erg\,s$^{-1}$                                                              & $\lesssim0.6$ & $<0.2$                                                                                              & $L_{\rm AGN}\sim10^{47}$erg\,s$^{-1}$   & {\bf 40} & $\approx1$ \\
AGN outflows    & $\lesssim200\,M_\odot$~yr$^{-1}$                                                                               & 0.3           & $\ll$0.1                                                                                            & {\bf 8000}~$M_\odot$~yr$^{-1}$          & {\bf 20} & $\approx1$? \\
SF outflows     & 1200 $M_\odot$~yr$^{-1}$                                                                                       & 1              & $<0.1$                                                                                            & 120~$M_\odot$~yr$^{-1}$                 & negl.    & negl. \\
SF photoionization       & 1200 $M_\odot$~yr$^{-1}$                                                                                       & 10              & $<0.03$                                                                                               & 120~$M_\odot$~yr$^{-1}$                 & 1      & $<1$ \\
Gravitational energy        & \begin{tabular}{@{}c@{}}$M_{\rm DM}=4\times10^{13}\,M_\odot$ \\ {\bf 10\,000}~$M_\odot$~yr$^{-1}$\end{tabular}
                & {\bf 160}                                                                                                      & $\approx1$ ?   & \begin{tabular}{@{}c@{}}$M_{\rm DM}=3\times10^{12}\,M_\odot$ \\ $500\,M_\odot$~yr$^{-1}$\end{tabular}
                & $\approx1$                                                                                                     & $<0.01$ ? \\
\hline
\hline
\end{tabular}\\
}
{{ Energy rates for \lya\ ionization. Total rates correspond to the whole energy rates available, while Effective estimates attempt to capture the fraction of all energy that can be used to power \lya\ emission (this by definition is capped to $\sim1$). The Constrain column details the key ingredient used to estimate these rates, as discussed in the text.  We express these energy rates}  in units of $1.5\times10^{44}$erg\,s$^{-1}$  for the case of outflows and gravity, i.e. the typical \lya\ luminosity of both the RO-1001 and also of the 
  QSO nebulae (Borisova et al. 2016), and relative to the required numbers of ionizing photons for the case of AGN/SF photoionization. The typical SFR of bright QSO fields is from  Schulze et al. 2019. {  Boldface values correspond to most energetically relevant entries, for each class.}
}
\end{table*}

\section{Discussion and interpretation}

{As typically the case with giant \lya\ nebulae, there are two fundamental questions that arise concerning their nature: 1) what is the origin of the cold \lya-emitting gas? and 2) what is the energy source of the emission?
In order to gather physical insights into these questions for the RO-1001 nebula, it is relevant to compare  with nebulae found in QSO/radio-galaxy fields} (Borisova et al. 2016), that are known to reach similar luminosities and maximum spatial extent. However, in RO-1001 there is no evidence of ongoing (obscured or not) AGN  activity from ultra-deep Chandra and mid-IR constraints (see Sect.~2.6).  
A summary of relevant physical quantities and energetics comparing the RO-1001 nebula and QSO nebulae is reported in Table~3.  { We would like to reiterate that a key ingredient in this comparison is the assessment of the hosting dark matter halo mass, which is presented in this paper for RO-1001 and available statistically for the average QSO. The lack of this crucial information for Lyman Alpha Blobs from the literature prevents us from extending the same comparison to those objects.}  

\subsection{Origin of the cold gas: inflow and outflow rates.}

Regarding question 1) on the origin of the cold, diffuse, intra-group gas shining in \lya, there are two main primary channels that need to be evaluated and, ideally, distinguished: cosmological inflows, mainly expected to be a function of the DM halo mass and redshift, and outflows from AGNs/galaxies, which depend on the AGN bolometric luminosity  and galaxy SFRs. Secondary channels are also important to evaluate, namely the returning gas from outflows in the form of galactic fountains, and the cold gas that results from cooling due to {  hydrodynamic instabilities in the intra-group medium arising from the interaction between rapidly moving outflows/inflows and the hot gas}. However, they are directly related to the primary channels, given that e.g. the amount of hydrodynamic cooling that results from outflows (e.g., Qiu et al. 2020) largely depends on the outflow rates and the same is true for inflows. Similarly, we would ascribe the galactic fountains as mainly related to outflows (even if the gas is infalling).  Hence we focus the discussion on the primary channels. 

For the case of the QSO nebulae, outflow rates driven by AGNs (derived using the scaling relations in Fiore et al. 2017 based on the average bolometric luminosity of $10^{47}$~erg~s$^{-1}$) are expected to exceed inflow rates by at least two orders of magnitude (Table~3). This makes efforts to reveal infalling gas in those systems particularly challenging: the putative observational features of any infalling gas would have to be distinguished from those of the vastly larger amount of  gas originated from outflows (and at least in part returning in the form of fountains). 
The QSO-fields inflow rates are conservatively (over-)estimated using a maximally large hosting DM halo for QSOs  of $3\times10^{12}M_\odot$, where the best estimate from clustering is instead  $1\times10^{12}M_\odot$ (see Pezzulli et al. 2019).

The situation is reversed for the case of RO-1001: given the AGN luminosity upper limits and ongoing SFRs, infalling gas rates are expected to exceed outflow rates by  one order of magnitude. Most of the outflowing gas mass is expected to be SFR driven rather than AGN driven, see Table~3. For SF-driven outflows we assume a loading factor of 1, with outflow rates from star-formation equal to SFRs, plausible given the large stellar masses (Hopkins et al. 2012;
Newman et al. 2012; Gabor \& Bournaud 2014, Hayward et al. 2015). 

Hence, in relative terms, for the RO-1001 group, with respect to QSO fields,  { we estimate} a 3 orders of magnitude (a factor of 1000) higher ratio  between gas mass inflow and outflow rates { (roughly speaking, in QSOs the ratio of inflow/outflow rates is expected to be of order $10^{-2}$, while in RO-1001 it is of order 10).}
On the other hand, the larger hosting DM halo of RO-1001 will result in a higher proportion of the outflowing gas being retained in the system's deeper potential well and being recycled onto the galaxies via galactic fountains. This likely reduces somewhat the expected order of magnitude contrast between nearly pristine cosmological inflowing gas and material ejected from member galaxies via outflows (see, e.g., Valentino et al. 2016)\footnote{This is equivalent to saying that the outflowing gas would be seen when outflowing, and when part of it will be recycled back to the galaxies, implying that the outflowing contributions listed in Table~3 would have to be counted with a factor slightly above 1 (but well less than 2).}.
Similarly, not all the predicted cosmological inflows might remain cold while penetrating the halo, as discussed in the previous section, although instabilities from the flowing gas will generate secondary cold gas within the halo. Finally, when considering that the gas might be accumulating over longer timescales, the impact of AGN outflows in QSO fields might be somewhat reduced, noting that the survival time of the cold gas (Klein et al. 1994; Valentino et al. 2016; Schneider \& Robertson 2017) is not necessarily much longer than QSO variation timescales (both of order of 10~Myr).  This point is still debated, and longer gas survival times might be possible (Gronke \& Oh 2018; Mandelker et al. 2019). 
{ Nonetheless, the higher impact of accretion relative to outflows in RO-1001 over QSO fields is likely to remain significant. }

\subsection{Energetics}

Regarding question 2) concerning the powering source,
relevant channels are: ionization from AGN or star formation, dissipation of kinetic energy carried out either by outflows from the same AGN/SFR or from inflows from the cosmic web, ultimately due to gravitational energy. 

\subsubsection{AGN photoionization}

For the case of  photoionization, the observed \lya\ luminosity in RO-1001 corresponds to $\sim1.5\times10^{55}$ hydrogen ionizing photons. Scaling from CL~1449 calculations (Valentino et al. 2016), this ionizing photon rate requires an AGN with bolometric luminosity of L$_{\rm AGN}\sim$3.5$\times$10$^{45}$~erg~s$^{-1}$. In the RO-1001 nebula our limit on AGN activity (or, equivalently, assuming ongoing AGN activity at the cosmic average given the stellar mass and SFR present; Delvecchio et al. 2020), implies that $<60$\% of the required photons are produced. When considering that the typical Lyman continuum escape for moderate AGNs is $\sim30$\% { (reducing the effective energy to 20\% of that needed to ionize the nebula)}, and that not all photons will likely power the \lya\ nebula due to geometry constraints, covering factors, etc., we conclude that AGN ionization cannot produce the observed \lya. As discussed in Sect.~2.6, the possibility that a bright QSO just switched off seems very unlikely: the probability of finding a luminous QSO  associated with our massive galaxies  is $<10^{-4}$. Finally, a skeleton with three filaments onto a giant \lya\ nebula is hardly compatible with a spherical or conical geometry generally associated with QSO illumination.

For the QSO nebulae, \lya\ emission is generally assumed to be powered by  QSO ionization  and subsequent recombination, which indeed is the most energetic source (Table~3) given the average QSO luminosity of $10^{47}$~erg~s$^{-1}$, providing 40 times more ionizing photons than required to account for the typical \lya\ nebulae in these environments. 

\subsubsection{Energy injected from AGN and SF outflows}

We compute luminosity rates that can be induced by outflows as their mass flows times their typical velocities squared. Calculation of gas mass outflow rates for QSOs and for the RO-1001 nebula have been reported in Sect.~4.1 and Table~3. 
We  assume typical velocities of 500~km~s$^{-1}$ (1000~km~s$^{-1}$) for SF (AGN) outflows.

For the RO-1001 group the  SF-driven  outflows (with a smaller, possible contribution from AGN outflows) would carry out of the galaxies  just as much energy as that observed in \lya. However,  most of it will be  dissipated via thermal instabilities and only a (small) fraction of this energy would emerge reprocessed by \lya\ (Valentino et al. 2016; { effective energies lower by an order of magnitude with respect to total, see also Table~3)}. Also, their modest velocities might prevent them from reaching large distances from the ALMA galaxies (which contain most SFR in the RO-1001 structure). Hence we expect that outflows could contribute locally, but likely not dominate.

For QSOs, outflows carry 20 times the energy required to power the typical observed \lya\ luminosities and with much higher velocities than star formation driven outflows, thus possibly far reaching. They might thus contribute a sizeable amount of the observed \lya\ energy when dissipating  in the circumgalactic medium (see also Ji et al. 2019; Gronke \& Oh 2020a; Fielding et al. 2020). Even accounting for substantial inefficiencies in \lya\ production (as discussed previously for RO-1001), their role in contributing to the powering of QSO nebulae might have been underestimated so far (Table~3).  

\subsubsection{Ionization from star forming galaxies}

While this channel is negligible for QSOs \lya\ nebulae (when compared to the QSO ionization), for RO-1001
the output from the ongoing SFR from the three ALMA galaxies is potentially capable of producing 10 times more ionizing photons than are needed to ionize the \lya\ nebula. 
This channel must therefore be carefully investigated for the RO-1001 group. 

The ALMA galaxies are highly obscured. Even assuming  typical attenuation properties of normal main sequence galaxies of the same stellar mass (Pannella et al. 2015) would imply that only a  small fraction of  ionizing photons and \lya\ photons can actually escape such galaxies, on average. This number could be even smaller if these sources are obscured like typical SMGs (Simpson et al. 2017; Jin et al. 2019; Calabro et al. 2018; 2019), as IR-luminous massive galaxies at high $z$ tend to be (Elbaz et al. 2018; Puglisi et al. 2019). 
We emphasize that, unlike AGN obscurations, dust attenuation in star-forming galaxies is not expected to be highly anisotropic in the UV (it is not driven by a torus), as demonstrated, e.g., by the tight relation  between dust extinction and stellar mass (Garn \& Best 2010; Kashino et al. 2013). Hence we estimate what fraction of the ionizing photons (or \lya\ photons) can escape the SF galaxies in RO-1001 by, first, evaluating the UV rest-frame output of these sources towards our  direction.
The brightest object in the UV is galaxy 'C' (Fig.~7), that also has some  weak \lya\ enhancement at its position (Fig.~2). { We estimated its  UV SFR from the observed flux at 1500\AA\ rest-frame, derived from the photometry in the V and R-bands, and using the conversion from Daddi et al. 2004. This returns an estimate of $2.5M_\odot$~yr$^{-1}$ at the level of 1\% of  the IR SFR for this galaxy} (that contains 20\% of the total IR from the group). Galaxies A and B have much less than 1\% of the intrinsic UV radiation (as inferred from the IR) being emitted in the UV after escaping dust extinction. So in total the emerging, unattenuated UV SFR corresponds to  $<3\times10^{-3}$ of the SFR$=1200M_\odot$~yr$^{-1}$ seen in the IR, corresponding already to a  negligible $<3$\% fraction of the photons required  to power the observed \lya\ nebula.
The overall effective output from SF galaxies would hardly change when considering additional UV-selected galaxies in the RO-1001 overdensity (Fig.~7; notice that additional UV-bright sources aligned with the nebula as visible in Figs~2 and~7 are in the foreground). 

Still, further reductions have to be considered. For example, to predict the emerging \lya\ flux from the ongoing SFR one has to include additional differential attenuation between the UV rest frame and \lya.  It has been shown locally that in ULIRGs the relative escape fraction of \lya\ with respect to far-UV unattenuated regions is typically  at the level of 0.1\%, altough reaching 10\% in one peculiar case (Martin et al. 2015b). Hence we can expect \lya\ directly produced by the SFR in RO-1001 to be entirely negligible, including \lya\ photons directly produced at the sites of the ALMA galaxies and any contribution from scattering out of these sources.
In terms of computing the Lyman continuum escape from the SFR in RO-1001, required to ionize hydrogen and produce \lya\ by recombination, one has to further account for additional extinction between \lya\ and wavelengths below the Lyman break, including absorption by the HI gas within the galaxies. 
 
Finally, one can wonder if a large,  additional population of faint low-mass star forming galaxies, unseen to current limits and potentially attenuation-free, could produce the observed \lya.
Using deep Subaru Suprime-CAM imaging in VRI bands places a $5\sigma$ lower limit of $EW>500$ \AA\ (rest-frame) on the diffuse \lya\ equivalent width, using regions where no detectable continuum is present in the current UV rest frame imaging. This is high enough to exclude diffuse in-situ star formation (e.g., \lya\ originating in large numbers of low-mass, star-forming galaxies). { The consistency between the integrated flux from ALMA sources with the SCUBA2 total flux also supports this conclusion, leaving little room for any further diffuse SFR component.}

\subsubsection{Gravitational energy.} 

For the dark matter halo estimated for RO-1001, the  energy associated (Faucher-Gigu\`ere et al. 2010) with cosmological gas accretion is 160$\times$ what is required to power the nebula. 
This exceeds all other sources of energy by about two orders of magnitude (Table~3).  In the case of QSOs the same calculations show that gravitational energy connected with gas accretion is barely comparable to the energy required to  power the nebulae, and 100 times smaller than the energy that can be produced in \lya\ by QSO photoionization, hence it plays quite clearly a negligible role given the overall poor efficiency of converting such energy into \lya\ photons. 

In a RO-1001-like halo, the emerging \lya\ radiation powered by gravitational energy has been predicted by existing models (Dijkstra et al. 2009; Goerdt et al. 2010; Rosdahl \& Blaizot 2012; Laursen et al. 2019) to be of the order of $10^{44}$erg~s$^{-1}$, fully consistent with what is observed here, and at the level of few percent of the total available gravitational energy. On the other hand, predictions for the much smaller QSO halos from the same studies return average \lya\ luminosities $\sim 10$ times smaller than what is observed, on average. 

{ It is worth noting that giant radio-galaxies are typically hosted in much more massive halos than luminous QSOs (see, e.g., Nusser \& Tiwari 2015), hence they might behave more similarly to RO-1001 in terms of energetics, with proportionally larger contributions from gravitational energy associated to accreting gas. Albeit the presence of the bright radio AGN complicates the observational interpretation of those systems, evidence for \lya\ powered by streams has been discussed by Vernet et al. 2017 for  MRC~0316-257 at $z=3.12$, and a remarkable case of accretion of molecular gas has  been presented  by Emonts et al. 2021 for  4C~41.17 at $z=3.792$.}

\subsubsection{Cooling from the X-ray gas}

Localized runaway cooling might occur in the densest regions of the hot X-ray halo
due to the onset of thermal instabilities (Gaspari et al. 2012, Sharma et al. 2012).
The thermal cascade would then result in the emission of Lya photons which might contribute
on scales smaller than our resolution. This mechanism is the basis for the self-regulated 
feedback successfully explaining several observed features of the ionized filaments in local 
cool-core clusters (Voit et al. 2017). However, this mechanism is unlikely to explain the extreme \lya/X-ray luminosity ratio
of the giant nebula in RO1001 and similar objects, considering both a classical stationary
cooling flow (Geach et al. 2009) or empirically comparing to the observed values  \lya/X-ray ratios of cool-core clusters that are orders of magnitude lower than what we observe  in RO-1001 (see also discussions about this point in Valentino et al. 2016, for Cl1449 at $z=1.99$).
 
\subsubsection{Conclusions on the energetics}

In conclusion, energetic arguments strongly favor cooling via radiation-dissipated gravitational energy as the most plausible channel for the (collisional or shock)
 excitation of \lya\ emission in  RO-1001 nebula.  This channel is providing two orders of magnitude more energy than any other plausible channel for RO-1001. 
 This is in contrast to QSO nebulae, where the combined effect of QSO photoionization and outflows provides two orders of magnitude more energy than gravitation.
 Hence this results in a contrast of 4 orders of magnitude (factor of $10^4$) in relative terms between the RO-1001 and QSO fields in terms of likelihood of revealing \lya\ powered ultimately by gravitational energy release.

\subsection{The origin of \lya\ in RO-1001}

In light of these results it is relevant to re-evaluate which process might be responsible for the \lya\ emission. The classic expectation would be collisionally excited \lya\ from the cold gas dissipating kinetic energy acquired via the gravitational energy. In order to be viable this would require a non negligible neutral fraction in the gas. The formal HI column density inferred from \lya\ modeling (Sect.3.1; Fig.11) is fairly low, and not obvious to be sufficient for the purpose. For example, assuming the rough calculation of flowing gas mass reported in Sect.~3.1 and using the virial velocity and virial radius to compute the timescale corresponds to a  fraction of  neutral hydrogen mass over total gas mass of order  $10^{-3}$. This is still compatible with collisional excitation without the need to advocate photo-ionization from currently unknown sources, if the cold gas temperature is a few $10^4$~K (e.g., see Fig.1 in Cantalupo et al. 2008).  All the more, it should be emphasized that the HI column density and infall/outflow velocities inferred from the shell modelling reported in Sect.3.1 and Fig.~11 must be considered as strict lower limits. This is because the emerging spectrum is strongly weighted by sightlines with the lowest column densities, given that \lya\ escapes through the paths of least resistance (e.g., Eide et al. 2018), generally orthogonal to the stream velocities in case of infall (Gronke et al. 2017). 

We notice though that an alternative scenario suggested by recent modeling is that radiation is emitted through the cooling of mixed gas occurring at the boundary between the phases (Mandelker et al. 2019; Cornuault et al. 2018). In that case \lya\ radiation would originate from the combined dissipation of kinetic energy of the stream and of thermal energy from the mixed gas that cools down (Gronke \& Oh 2020a). 

This scenario, and our calculations of gas mass flows in Sect.~3.1 also imply 
 a total cold gas mass  present at each moment in the diffuse streams of order of $10^{11}M_\odot$ (to be compared to lower limit of $\approx10^{8}M_\odot$ of neutral hydrogen). Is that a reasonable gas mass to be diffused in the intra-group gas? 
 On one hand this is comparable or smaller than the ISM mass of all group galaxies combined, and less than 10\% of all baryons expected to be in the group given its DM estimate and assuming a universal baryon fraction. Assuming pressure equilibrium the flowing gas would need to be confined in clouds with quite small volume filling factor, of order of $10^{-3}$ to $10^{-5}$, qualitatively similar to what discussed in theory work (e.g., Cornuault et al. 2018; McCourt et al. 2018; Gronke \& Oh 2020b). This is required to reach a high enough density to compensate for the much larger temperature of the hot medium. Such dense gas clouds would produce copious recombination \lya\ photons following photo-ionization, if a source emitting enough ionizing photons towards this medium were to be present.  This does not appear to be the case for RO-1001, at least given present evidence.
 
\section{Summary and conclusions}

The main findings of this work can be summarized as follows:

  \begin{enumerate}
      \item RO-1001 is a group at $z=2.91$ defined by 4 galaxies with stellar masses above $10^{11}M_\odot$, originally selected as an overdensity of faint radio sources (12$\sigma$ excess) over a small,  10$''$ region.
      \item It contains an ongoing SFR of 1200~M$_\odot$~yr$^{-1}$ mostly spread among 3 ALMA detected, IR-luminous sources characterized by very compact IR sizes, for which we reported CO[3-2] detections with NOEMA.
      \item RO-1001 does not contain any detectable evidence of ongoing AGN activity, down to limits that are consistent with a cosmic average co-existence of AGN and SFR at typical levels.
      \item The hosting dark matter halo mass is estimated to be typical of a group, with $\sim4\times10^{13}M_\odot$ consistently derived using three methods, including a blind X-ray measurement with significance of 3.1$\sigma$.
      \item RO-1001 hosts a giant \lya\ halo with a luminosity of $1.3\times10^{44}$~erg~s$^{-1}$ as revealed by KCWI observations. Three \lya\ filaments are observed extended over an overall 300~kpc area (comparable to the virial diameter of RO-1001) and converging into the bright \lya\ core whose luminosity peak is well aligned with the center of mass of the group.
      \item The \lya\ emission is fairly 'cold', with a velocity dispersion of about 250~km~s$^{-1}$. The absolute velocity at the edge of the filaments is of order of 400-500~km~s$^{-1}$ higher than in the core, comparable to the virial velocity. 
      \item The \lya\ spectral profile in most of the core appears to be dominated by blueshifted components, as well as in the two main filaments as derived by multi-Gaussian decompositions. Evidence for regions dominated by redshifted components in \lya\ exists as well.
      \item Shell modeling of the blueshifted emission in the core suggests the presence of inflowing gas with moderate velocity (150~km~s$^{-1}$) and column densities of neutral gas (a few $10^{17}$~cm$^{-2}$) arising from a rest-frame velocity consistent with the average one as traced by CO in the ALMA galaxies. This supports the \lya\ moments analysis as being relatively unaffected by resonant scattering.
      \item From the point of view of expected rates of gas flows given the hosting DM halo masses and ongoing SF and AGN activity, cold gas inflows from the cosmic web are expected to dominate over outflows by up to an order of magnitude  in RO-1001. This corresponds to a relative contrast of 3 orders of magnitude with respect to \lya\ nebulae hosted by luminous QSO (where inflows from the cosmic web are expected to be two orders of magnitude smaller than outflows).
      \item From the point of view of \lya\ powering and energetics, the gravitational energy associated with the gas infall can provide two orders of magnitude more energy than required to power the observed \lya\ nebula in RO-1001, and over two orders of magnitude more energy than any other plausible source. Again, this corresponds to a relative contrast of 4 orders of magnitude with respect to QSO fields (where the \lya\ is powered by photo-ionisation and sub-sequent recombinations, with possibly a contribution from outflows).
      \item { A large range of observational properties of the RO-1001 \lya\ nebula are consistent with predictions from cold accretion models for halos of the same DM halo mass. This includes the overall luminosity, surface brightness levels, area, velocity and velocity dispersion, linear and transverse sizes of the filaments, and of course the {\em canonical} number of three filaments.  }
      
    \end{enumerate}
    
    In conclusion, RO-1001 at $z=2.91$  { provides a  plausible observation of gas accretion towards a massive potential well} (Fig.9), with its filaments possibly identifiable with cold accreation streams, but where the effect of phase mixing, dissipation and local cooling seems also important.  
Knowledge of the mass and position of the center of mass of the RO-1001 group is crucial information that was not available for other known filamentary nebulae and sets a clear new precedent for future research along this line.
\smallskip 

Of course, several uncertainties remain. We do see evidence that even in RO-1001 outflowing gas is still playing some non-negligible role, and it is extremely difficult to  definitely rule out photo-ionization as the dominant mechanism for \lya\ emission. More insights could be obtained if we were to be able to obtain measurements of non resonant lines that, as discussed in Sect.3.2, would allow a more solid modeling of the \lya\ emission  in terms of unveiling the supposed prevalence of inflows while at the same time providing more robust kinematics and velocity dispersion fields. Observations of H$\alpha$ with JWST could be illuminating, also keeping in mind that fairly weak H$\alpha$ emission would be expected if \lya\ is predominantly collisionally excited. Similarly, UV metal line observations would provide constraints on ionization and enrichment, potentially clarifying if we are seeing fairly pristine gas being accreted, e.g. at least at the edge of the filaments. All of this will have to wait for future follow up of RO-1001 and other structures in coming years. 

It is not obvious that trying to observe \lya\ around other (non-QSO) structures hosted in lower mass  halos would provide an advantage in terms of finding  more convincing probes of infalling gas then what we could gather so far in RO-1001. In relative terms, the higher fraction of the inflowing gas that could remain cold after entering lower mass halos would be counterbalanced by the lower contrast between infall and outflows owing to the different scaling of these terms with mass and to the increased loading factor of outflows from lower mass galaxies. In absolute terms, the \lya\ luminosities of infalling gas would be suppressed, roughly proportionally to the hosting halo mass, according to  model predictions. Of course, it is also not obvious that model predictions are, even roughly, correct. We have emphasized how predictions of \lya\ emission from cold streams for  a RO-1001-like halo are very close to the observed \lya\ luminosity. On the other hand, models might be simplistic.  We already recalled that BCG assembly happens too early in the models, compared to observations, affecting in some way how gas is fueled to galaxies in the inner halos. {Given that cold streams are driven by gravity, that in turn are mostly affected by the total dark matter distribution rather than the location of individual galaxies, this discrepancy would likely have little effect on the reliability of cold accretion predictions -  at least on relatively large scales. However, it might reflect issues on how the accreting gas is effectively fueling galaxies at the center of massive halos at high redshifts. }
More specifically in terms of \lya, Faucher-Gigu\`ere et al. (2010) suggested that  accounting for self-shielding and  properly treating sub-resolution effects might easily lead to reduced forecasts of the \lya\ emission by 1 or more orders of magnitude. Rosdahl \& Blaizot (2012) predict a substantially higher neutral fraction than is inferred from our shell modeling of the emerging spectrum. If this were to be true (but beware that column density from shell modeling are strict lower limits to real average column densities, see comments in Sect.~4.3), than  their calculations of emerging \lya\ emission might also be overestimated, given that collisional excitation luminosity scales with neutral gas density. Needless to say, if models are optimistic by large factors then it might become prohibitive to ever detect any signature of cold accretion from \lya\ emission. Ultimately, future generations of models capturing physical effects that currently remain sub-resolution will be crucial for interpreting this and future observations in terms of cold gas accretion. 

{ The environments of QSOs have provided a remarkable and diverse population of giant \lya\ nebulae, but due to the presence of the photoionizing quasar, it will  be prohibitively difficult to use those to securely argue for the detection of a cold cosmological flow.  In this paper, we show that follow-up of massive high redshift groups/clusters (e.g., as in this case from radio-based selection of overdensities) can lead to the discovery of a new class of giant \lya\ nebulae, whose emission is plausibly powered by the collisional ionization of a cosmological cold flow.  Additional studies of giant \lya\ nebulae centered of massive high-redshift dark matter halos will provide new constraints on this process, that may be important in fueling the rise of giant clusters and massive galaxies.}

\begin{acknowledgements}
We are indebted to Sebastiano Cantalupo for use of his CubeEx software and for enlightening discussions. We also thank Matt Lehnert, Avishai Dekel, Nir Mandelker, Anne Verhamme and Mauro Giavalisco for discussions. This work includes observations carried out with the IRAM NOEMA Interferometer. F.V. acknowledges support from the Carlsberg Foundation Research Grant CF18-0388 "Galaxies: Rise And Death" and the Cosmic Dawn Center of Excellence funded by the Danish National Research Foundation under then Grant No. 140. RMR acknowledges financial support from HST-GO-15910.  MG was supported by NASA through the NASA Hubble Fellowship grant HST-HF2-51409 and acknowledges support from HST grants HST-GO-15643.017-A, HST-AR-15039.003-A, and XSEDE grant TG-AST180036. AP acknowledges financial support from STFC through grants ST/T000244/1 and ST/P000541/1. S.J. acknowledges financial support from the Spanish Ministry of Science, Innovation and Universities (MICIU) under AYA2017-84061-P, co-financed by FEDER (European Regional Development Funds). VS acknowledges support from the ERC-StG ClustersXCosmo
grant agreement 716762.
 IRAM is supported by INSU/CNRS (France), MPG (Germany) and IGN (Spain). This paper makes use of data from ALMA: a partnership of ESO (representing its member states), NSF (USA) and NINS (Japan), together with NRC (Canada), MOST and ASIAA (Taiwan), and KASI (Republic of Korea), in cooperation with the Republic of Chile. The Joint ALMA Observatory is operated by ESO, AUI/NRAO and NAOJ. The National Radio Astronomy Observatory is a facility of the National Science Foundation operated under cooperative agreement by Associated Universities, Inc.  This work was supported by the Programme National Cosmology et Galaxies (PNCG) of CNRS/INSU with INP and IN2P3, co-funded by CEA and CNES.
\end{acknowledgements}

  \begin{appendix}
  
  \section{SNR ratio per pixel}
  
  We show in Figure A1 the SNR of the \lya\ detection in the original, unsmoothed map. 
 
 \begin{figure*}
{\centering  
\includegraphics[width=0.89\textwidth, angle=0]{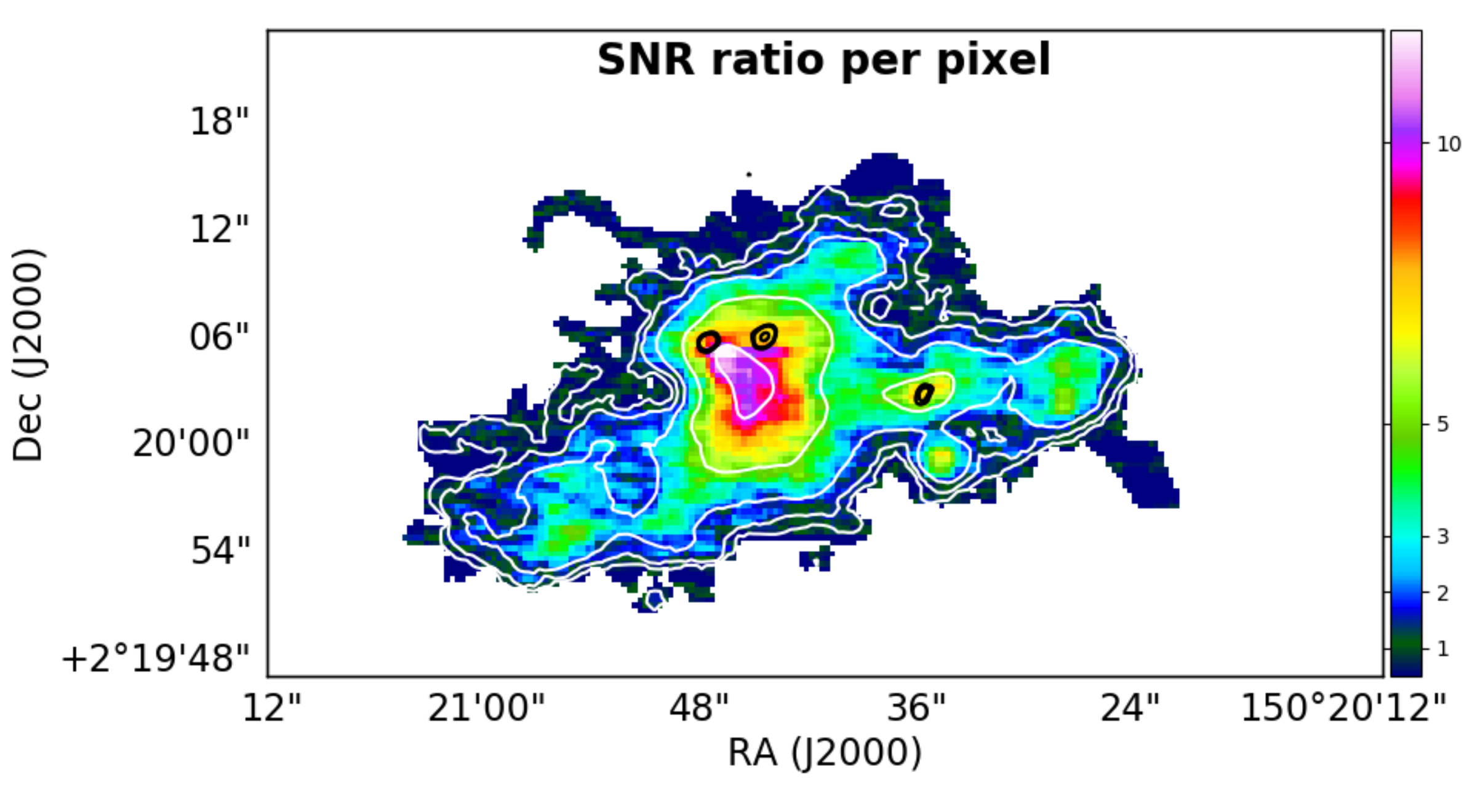}
}
\caption{
The color coding in the image (right scale) shows the SNR ratio per pixel in the original, un-smoothed data, over the region of the RO-1001 nebula within the area where a detection is found in the adaptively smoothing map. The flux error in each pixel is already enlarged to account for correlated noise (see Section 2).
ALMA continuum sources  are shown as black contours.
White contours show the \lya\ surface brightness levels from Fig.~2.  }
\end{figure*}

 \end{appendix}

%
%

 \bibliographystyle{aa}

\end{document}